\documentclass[12pt]{iopart}

\usepackage{color}
\usepackage{graphicx}  
\expandafter\let\csname equation*\endcsname\relax
\expandafter\let\csname endequation*\endcsname\relax
\usepackage{amsmath}
\usepackage{amssymb}
\usepackage{times}
\usepackage{bm}
\usepackage{acronym}
\usepackage{amsfonts, hyperref}
\def\be{\begin{equation}}
\def\ee{\end{equation}}

\def\bea{\begin{eqnarray}}
\def\eea{\end{eqnarray}}

\begin{document}
\title[Observing and measuring the neutron-star equation-of-state]{Observing and measuring the neutron-star equation-of-state in
spinning binary neutron star systems}

\author{Ian Harry}
%\email{ian.harry@aei.mpg.de}
\address{Max Planck Institute for Gravitational Physics (Albert Einstein Institute),
  Am M\"uhlenberg 1, D-14476 Potsdam-Golm, Germany}
  
\author{Tanja Hinderer}
%\email{hinderer@science.ru.nl}
%\address{Max Planck Institute for Gravitational Physics (Albert Einstein Institute),
%  Am M\"uhlenberg 1, D-14476 Potsdam-Golm, Germany}
\address{Department of Astrophysics/IMAPP, Radboud University, P.O. Box 9010,
6500 GL Nijmegen, The Netherlands}

\begin{abstract}
LIGO and Virgo recently made the first observation of a binary neutron
star merger demonstrating that gravitational-wave observations offer
the ability to probe how matter behaves in one of the
most extreme environments in the Universe.
However, the
gravitational-wave signal emitted by an inspiraling binary neutron star
system is only weakly dependent on the equation of state and extracting
this information is challenging.
Previous studies have focused mainly on
binary systems where the neutron stars are spinning slowly and the main imprint of neutron star
matter in the inspiral signal is due to tidal effects.
For binaries with non-negligible neutron-star spin the deformation of the neutron star
due to its own rotation introduces additional variations in the emitted gravitational-wave signal.
Here we explore whether highly spinning binary neutron-star
systems offer a better chance to measure the equation-of-state
than weakly spinning binary-neutron star systems.
We focus on the dominant adiabatic
quadrupolar effects and consider three main questions.
First, we show that equation-of-state effects
can be significant in the inspiral waveforms, and that the spin-quadrupole effect dominates for
rapidly rotating neutron stars.
Second, we show that variations in the spin-quadrupole
phasing are strongly degenerate with changes in the component masses and spins, and therefore neglecting these terms has a
negligible impact on the number of observations with second generation observatories.
Finally, we explore the bias in the masses and spins that would be introduced by using
incorrect equation-of-state terms.
Using a novel method to rapidly evaluate an approximation of the likelihood we show that assuming
the incorrect equation-of-state when measuring source parameters can lead to a significant bias.
We also find that the ability to measure the equation-of-state is improved when considering spinning
systems.
\end{abstract}

%\keywords{Binary neutron stars, equation-of-state, spin-quadropole}
%\submitto{\cqg}

\maketitle

\section{Introduction}
\label{sec:intro}

On August 17, 2017, Advanced LIGO~\cite{TheLIGOScientific:2014jea} and Advanced Virgo~\cite{TheVirgo:2014hva}
made the first observation of a binary neutron-star merger~\cite{TheLIGOScientific:2017qsa}.
This merger was associated with the short gamma-ray burst GRB170817A~\cite{GBM:2017lvd} and prompted
a large-scale observing campaign to characterize the accompanying electromagnetic transients from the
entire electromagnetic spectrum~\cite{GBM:2017lvd}. This observation firmly established the field of
multimessenger astronomy, and demonstrated its potential to directly probe the
physics of neutron stars. Neutron stars are exceptional environments where all four fundamental forces
are simultaneously important and consist of matter compressed by their strong self-gravity to densities
up to several times the density of an atomic nucleus. Despite much recent progress in theory, experiments,
and observations, see
e.g.~\cite{Danielewicz:2002pu,Tews:2012fj,Kurkela:2009gj, Steiner:2010fz,Steiner:2012xt, Hebeler:2010jx,
Hebeler:2013nza,Horowitz:2000xj,Ozel:2010fw, Lattimer:2015nhk,Demorest:2010bx,Antoniadis:2013pzd},
determining the composition and equation of state of neutron star matter remains a major objective
at the forefront of fundamental physics~\cite{longrange} and astrophysics~\cite{decadal,NICER}.
In the coming years,  gravitational wave detections of many more binaries involving one or two
neutron stars are expected, and will offer new ways to explore the internal structure of neutron stars. 

% Tidal stuff has already been looked at
One avenue for measuring the equation-of-state of matter in
neutron star binaries is to look for deviations in the emitted gravitational waveform due to tidal coupling between
the inspiraling neutron stars. The dominant effect is due to an adiabatic linear tidal interaction~\cite{Flanagan:2007ix}.
Unfortunately, this effect only becomes important when the stars are very close to
merger, since it scales as $\sim k_2(R/r)^{5}$, where $k_2$ is the tidal Love number, $R$ is the neutron star's radius
and $r$ the orbital separation. For neutron stars the gravitational-wave frequency $f\sim 2 f_{\rm orbit}\sim \sqrt{GM/r^3}/\pi$ at
which tidal effects become noticeable is $\gtrsim$ a few hundred Hz, where the stars are just a few orbits away
from coming into contact and the sensitivity of gravitational wave interferometers has begun to decrease. 
% What about QM term
% The QM term depends on spin and EOS, if spin is large EOS effects might be measurable at 2PN, much larger effect than tidal
% How does this term affect the evolution?
However, for spinning neutron stars there is another equation-of-state-dependent effect that scales as
$\sim {\cal Q }S^2/r^2$, where $S$ is the spin angular momentum and ${\cal Q}$ a dimensionless coefficient
characterizing the star's rotational deformation away from the value of black holes ${\cal Q}_{\rm BH}=1$.
This spin-quadrupole term, first described in~\cite{Poisson:1997ha}, arises because a spinning compact object has
a non-spherically symmetric mass distribution due to the rotational flattening at the poles, which distorts the gravitational field around the neutron star. This
distortion affects the orbital evolution and the gravitational wave emission.

% BNSs can get large spins
It is important to understand if these different effects of neutron star matter
can be measured when using a gravitational-wave observatory like Advanced LIGO. It
is also important to understand if neglecting such terms in current searches will reduce the number of
observations that might be made in the coming years. This is because binary inspiral searches are based on cross-correlating the data with a bank of waveform templates, and unmodeled physics can potentially lead to a loss of signals due to inadequate templates. A number of works have addressed these
topics~\cite{Read:2009yp,Hinderer:2009ca,Markakis:2010mp,Damour:2012yf,DelPozzo:2013ala,Read:2013zra,Wade:2014vqa,
Lackey:2014fwa,Agathos:2015uaa,Cullen:2017oaz}, but in most cases did not include the spin-quadrupole
term in the waveform~\footnote{The work~\cite{Agathos:2015uaa}
did consider the spin-quadrupole term, but the results presented there are largely orthogonal to the presentation here.}.
The main reason why the spin-quadrupole term has largely been neglected is because of the expectation that a residual birth-spin of a
neutron star would have decayed away long before
entering the band of interest for ground-based gravitational wave observatories~\cite{2004hpa..book.....L}.
If the neutron-stars are non-spinning (slowly spinning)
the quadrupole moment term is not present in (only a small contribution to) the gravitational wave signal
and one must therefore rely only on the tidal
effects to gain any information about the equation-of-state. However, it is possible that a neutron star
can be spun up in a process
known as ``recycling''. Recycled neutron stars have been observed as millisecond pulsars with spin frequencies
as large as $f_{\rm spin}=716$Hz \cite{Hessels:2006ze}. This translates into a dimensionless spin of
order $\chi=S/m^2=2\pi(c/G) f_{\rm spin}I/m^2\sim 0.4$ assuming the pulsar's mass and radius are
$m\sim 1.4M_\odot$ and $R\sim 12$km, and with a moment of inertia $I\sim 1.4\times 10^{45}{\rm g}\, {\rm cm}^2$,
where the value of $I$ and thus the inferred spin $\chi$ for a given rotation frequency depend on the equation
of state (note also that the mass of the millisecond pulsar is not known). Such rapidly rotating neutron stars
have not yet been observed in binary neutron star systems and it is not
clear if binary neutron star systems where at least one of the bodies has such large spins will exist.
One possibility is that they could form in dense stellar environments such as globular clusters or galactic centers
through dynamical interactions~\cite{Samsing:2016bqm}.
However, if such systems \emph{do}
exist it may be possible to learn about the neutron star's internal physics by measuring the effect of the
spin-quadrupole term, in addition to the tidal deformability term on the orbital evolution.
Conversely, neglecting this effect might lead to a bias in all measured parameters, and potentially a loss of detected signals. 

% We investigate faithfulness and then effectualness
In this work we investigate whether deviations due to the neutron stars' equation-of-state would be observable in
binary-neutron star systems observed with Advanced LIGO, allowing spins to be as large as $\chi\sim 0.4$. We particularly focus
on the effect that the spin-quadrupole term can have and the effects that neglecting this term might cause.

The organization of this paper is as follows. In section~\ref{sec:waveforms} we briefly review the waveform model
used in our study and the dominant effects of the internal structure of the neutron stars.
In section~\ref{sec:searchintro} we give a brief introduction of the data analysis techniques used in this article.
In section~\ref{sec:faithfulness} we investigate
the similarity between waveforms whose masses and spins are equal, but where the equation of state is allowed to vary.
This does not give a complete picture though, because it is possible that a system with one equation-of-state could be
modeled by systems with a different equation-of-state but also different masses and/or spins. Therefore, in section
\ref{sec:effectualness} we investigate the ``fitting-factor'' between spinning binary neutron star waveforms with different
equation-of-states to determine if changes in the equation-of-state can be hidden by changes in the system's intrinsic parameters.
In section~\ref{sec:param_bias} we use a new method to efficiently evaluate the marginalized posterior probability
distribution for the source parameters for four different example cases to determine the bias in masses and spins
when neglecting the equation-of-state effects or using wrong values, as well as the improvements in equation-of-state
measurements for rapidly spinning binaries. In section~\ref{sec:redshift} we briefly discuss how the redshift might be
measured from equation-of-state terms. Finally we conclude in section~\ref{sec:conclusion}. Unless otherwise
specified we will use geometric units $G=c=1$. 

\section{Waveform model including the quadrupolar spin and tidal deformations of neutron stars}
\label{sec:waveforms}

We begin by discussing how the gravitational-wave signal emitted during a
binary neutron-star merger would differ from that of a binary black-hole merger with otherwise
identical source properties. For binary-neutron star mergers, most of the information in gravitational-waves 
will come from the inspiral, and not the merger or post-merger,
because the merger signal is emitted at frequencies too high to be easily observed with
second generation gravitational wave observatories. Therefore we only consider the inspiral
of the two bodies here.

In this work we will consider three effects that distinguish binary-neutron star mergers from binary
black hole mergers: First, the effect
of the rotational deformation of the components in the case that the components' spins are
non-zero~\cite{Poisson:1997ha}. Second, the effect of the tidal deformation of the neutron-stars~\cite{Flanagan:2007ix}, and third
the effect of the merger frequency of the binary~\cite{Bernuzzi:2015rla}.
Since we are interested in the dominant effects we neglect higher multipoles beyond the 
quadrupole~\cite{Hinderer:2009ca}, dynamical tidal interactions~\cite{Xu:2017hqo,Lai:1993di,Ho:1998hq,Hinderer:2016eia,Kokkotas:1995xe},
gravitomagnetic tides~\cite{Damour:2009vw,Landry:2015cva}, spin-tidal couplings~\cite{Pani:2015nua,Landry:2017piv},
and the presence of a surface rather than an event horizon~\cite{Maselli:2017cmm}. We note that there are some scenarios that
predict more exotic, or more extreme deviations between a binary-neutron star merger
and a binary-black hole merger~\cite{Essick:2016tkn}, we do not consider such scenarios here. We also
note that the most prominent imprint of neutron star physics on the gravitational waves is in
the post-merger signal that can markedly differ from the ringdown of a binary-black hole
merger~\cite{Bauswein:2011tp,Bauswein:2012ya}. However, we do not consider
the post-merger signals in this work because they occur at frequencies too high to be observable with current facilities,
and because while simulations of the post-merger signal do exist, this epoch
is generically not yet well understood due to the complexity of the physics that becomes important.

\subsection{Approximate frequency-domain description of GW signals from binaries}

We will model the gravitational-wave signal emitted by two inspiralling neutron stars
using the Post-Newtonian 
approximation~\cite{Blanchet:2006zz}. Throughout this work we assume that the objects move on circular orbits 
and that their spins are collinear with the orbital angular momentum. 
In the post-Newtonian framework, the gravitational wave phase evolution is computed by imposing that
the power radiated in gravitational waves is balanced by the change in binding energy of the binary.
A number of different perturbative expansions, in powers of $v/c$ where $v\sim \sqrt{GM/r}$ is the orbital velocity,
can be used to compute the phase evolution given
the center-of-mass energy, currently known to $O(v/c)^8$
beyond the Newtonian result~\cite{Damour:2016abl,Marchand:2017pir}, and the 
gravitational-wave flux, for which $O(v/c)^7$ corrections to the quadrupole formula have been
computed~\cite{Blanchet:2001ax}. In this work we consider the 
frequency-domain ``TaylorF2'' model describing the $\ell=|m|=2$ spherical harmonic mode of the waveform; for an overview
of the different Post-Newtonian waveform approximants see e.g. Ref.~\cite{Damour:2000zb}. TaylorF2 waveforms can be 
expressed in analytic form as
\begin{equation}
 \tilde h(f) = A(f;\mathcal{M},D_L,\theta_x) e^{ - i \Psi(f;\lambda_i)}.
\end{equation}
Here $\tilde h(f)$ denotes the Fourier transform of $h(t)$, the time-domain
gravitational-wave strain, 
\begin{equation}
\mathcal{M}=(m_1 m_2)^{3/5}(m_1+m_2)^{-1/5} 
\end{equation}
denotes the chirp mass, $D_L$ the 
luminosity distance to the source, and $\theta_x$ describes the various 
orientation angles that only affect the
amplitude and overall phase of the observed gravitational
waveform~\cite{Allen:2005fk}. The phase $\Psi$ is computed in the stationary phase approximation by integrating 
\begin{equation}
\label{eq:ODEpsi}
\frac{d^2 \Psi}{d f^2}=-\frac{(dE/df)}{{\cal L}_{\rm GW}},
\end{equation}
where $E$ is the energy of the system and ${\cal L}_{\rm GW}$ is the gravitational-wave luminosity. The result of
solving Eq.~\eqref{eq:ODEpsi} perturbatively for small $f$ and using the Post-Newtonian equations is 
\begin{equation} \label{eq:phase_exp}
\Psi = 2 \pi f t_c - \phi_c(\theta_x) + 
\sum_{i = 0}^{7} \sum_{j = 0}^{1} \lambda_{i, j} f^{(i-5)/3} \log^j f,
\end{equation}
where $t_c$ is the coalescence time and $\phi_c$ is a constant phase offset.
The coefficients
$\lambda_{i,j}$ for nonspinning point-mass binaries are given in~\cite{Arun:2008kb}. The first term, $\lambda_{0,0}$,
depends only on the chirp mass ${\cal M}$ but higher order post-Newtonian corrections also involve the symmetric mass ratio 
\begin{equation}
\eta=m_1 m_2(m_1+m_2)^{-2}.
\end{equation}
  Spin effects first enter at $O(v/c)^3$ and are characterized at that order by the spin-orbit parameter
%M
%\begin{widetext}

%\begin{eqnarray}
%\label{eq:lambdas35}
%\lambda_{0,0} &=& \frac{3}{128} (\pi \mathcal{M} f_0)^{-5/3}, \\
%\lambda_{2,0} &=& \frac{5}{96 \eta^{2/5}} \left( \frac{743}{336} + \frac{11}{4} 
%\eta \right) (\pi \mathcal{M} f_0)^{-1}, \\
%\lambda_{3,0} &=& -\frac{3 \pi}{8 \eta^{3/5}} \left( 1- \frac{1}{4 \pi} \beta 
%\right) (\pi \mathcal{M} f_0)^{-2/3}, 
%\end{eqnarray}
%\begin{eqnarray}
%\lambda_{4,0} &=& \frac{15}{64 \eta^{4/5}} \left( \frac{3058673}{1016064} + 
%\frac{5429}{1008} \eta + \frac{617}{144} \eta^2 - \sigma \right) (\pi 
%\mathcal{M} f_0)^{-1/3} \\
%\lambda_{5,1} &=& \frac{3}{128 \eta} \left(\frac{38645 \pi}{756} - \frac{65 
%\pi}{9}\eta\right) \\
%\lambda_{6,0} &=& \frac{3}{128\eta^{6/5}} 
%\biggl(\frac{11583231236531}{4694215680} - \frac{640 \pi^2}{3} 
%          - \frac{6848}{21} \left( \gamma_E + \log 4 - \frac{1}{5} \log \eta + 
%\frac{1}{3} \log (\pi \mathcal{M} f_0) \right) \nonumber \\
%          &-& \frac{15737765635}{3048192}\eta + \frac{2255 \pi^2}{12}\eta + 
%\frac{76055}{1728}\eta^2
%          - \frac{127825}{1296}\eta^3 \biggr) (\pi \mathcal{M} f_0)^{1/3} \\
%\lambda_{6,1} &=& -\frac{1}{128 \eta^{6/5}}\frac{6848}{21} (\pi \mathcal{M} 
%f_0)^{1/3} \\
%\lambda_{7,0} &=& \frac{3}{128\eta^{7/5}} \left(\frac{77096675 \pi}{254016} + 
%\frac{378515 \pi}{1512}\eta - \frac{74045 \pi}{756}\eta^2\right) (\pi 
%\mathcal{M} f_0)^{2/3},
%\end{eqnarray}
%where $\gamma_E$ is the Euler gamma constant and $f_0$ a reference frequency. The parameters $\beta$ (the dominant spin-orbit 
%coupling term)
%and $\sigma$ (the dominant spin-spin coupling term) are given by
%
\begin{equation}
\label{eq:beta}
\beta = \frac{1}{12} \sum_{i=1}^2 \left[ 113 \left(\frac{m_i}{m_1 + 
m_2}\right)^2 + 75 \eta \right]
\bm{\hat{L}} \cdot \bm{\chi}_i .
\end{equation}
Here, $\bm{\hat{L}}$ is the unit vector in the direction of the orbital angular 
momentum and the $\chi_i$ are the dimensionless spin parameters of each object 
\begin{equation}
\bm{\chi}_i=\frac{\bm{S}_i}{m_i^2}, \label{eq:chidef}
\end{equation}
 where $\bm{S}$ denotes the spin angular momentum. At second post-Newtonian order, $O(v/c)^4$, spin-spin interactions
 start to influence the signal and are parameterized by
\begin{equation}
\sigma =\frac{\eta}{48} \left( -247 \bm{\chi}_1 \cdot \bm{\chi}_2 + 721 
\bm{\hat{L}}
\cdot \bm{\chi}_1 \bm{\hat{L}} \cdot \bm{\chi}_2 \right).
\end{equation}

Spin-dependent contributions to the phase appear again at higher orders in $v/c$. The explicit results for
these contributions, currently known up to $O(v/c)^7$, can be found e.g. in \cite{Arun:2008kb}.

\subsection{Equation-of-state effects in the gravitational-wave signals}

For the purpose of our study we will consider two physical effects that lead to an imprint
of the equation-of-state on the gravitational waves: spin- and tidally-induced deformations. For each of these we will focus
on the dominant quadrupolar effect.
%
% SOME PN EQUATION maybe a couple of terms.
%
As discussed above, the spin of the objects first enters equation~\eqref{eq:phase_exp} as an order $(v/c)^{3}$
correction to the leading order $\lambda_{0,0}$-term, due to the coupling between the orbital angular momentum
and the components' spin encoded in $\beta$ given in equation~\eqref{eq:beta}. However, this term does not depend
on the deformation of the objects and is thus
independent of the nature of the object when expressed in this way~\footnote{This is different
from the context of radio observations of binary pulsars, where the spin-orbit effect in the
periastron advance is central in attempts to measure the equation-of-state-dependent moment of inertia $I$.
The reason is that for pulsar observations the measurable quantity is the spin period $P$,
which is used to replace $S=2\pi I/P$ in the spin-orbit couplings}. 

Finite size effects that depend on the underlying equation-of-state first enter the gravitational-wave signal as an
order $(v/c)^4$ correction through the quadrupole-monopole interaction. This effect
arises because a rotating neutron star is not spherically symmetric, which is physically
a purely Newtonian effect despite the Post-Newtonian-like scaling with the frequency. The star's rotation causes
a centrifugal flattening of its mass distribution into an oblate shape, which in turn
creates a distortion in the gravitational field it generates. At large distances from the star,
the leading order deviation of the Newtonian gravitational potential away from that of a
nonspinning body is characterized by a quadrupole moment scalar
\begin{equation}
{\rm Q}_{\rm spin}\approx - {\cal Q}({\rm m, EOS}) \chi^2 m^3,
\end{equation}
 where $\chi=|\bm{\chi}|$ is the
magnitude of the dimensionless spin defined in \eqref{eq:chidef} and ${\cal Q}$ is a
dimensionless parameter characterizing the quadrupole deformation~\cite{Laarakkers:1997hb,Poisson:1997ha}.
These Newtonian notions can be generalized to general relativity by considering the spacetime
of a rotating neutron star and identifying the quadrupole moment from the asymptotic fall-off
behavior of the metric potentials at large distances from the neutron star, or equivalently
by using a more formal definition of multipole moments \cite{Laarakkers:1997hb}. For a black hole,
the quadrupole is given by the exact relation ${\rm Q}_{\rm BH}= - \chi^2 m^3$, in
accordance with the no-hair property. The difference between the black hole and neutron star
quadrupole moments affects the orbital motion and rate of inspiral. This results in the
following leading order contribution to the phase 
%
%% SHow the 2PN term here
%
\begin{equation}
\lambda_{4,0}^{\text{ QM}}=\frac{30}{128\eta^{4/5} }\sigma_{\text{QM}} (\pi \mathcal{M} f_0)^{-1/3}, 
\end{equation}
where
\begin{equation}
\sigma_{\text{ QM}} =-\frac{5}{2}\sum_{i=1}^2{\cal Q}_i \chi_i^2\frac{m_i^2}{M^2}\left[3(\bm{\hat{\chi}}_i \cdot \bm{\hat{L}})^2-1\right].
\end{equation}
The spin-induced deformation term enters again at higher orders. In our analysis we include the $O(v/c)^6$ term that is given for the case where both spins aligned with $\hat L$ by ~\cite{Krishnendu:2017shb}
\begin{eqnarray}
\lambda_{6,0}^{\text{ QM}}&=&\frac{(\pi \mathcal{M} f_0)^{1/3}}{128 \eta^{6/5}}\left[\frac{2215\eta^2}{2(1-2\eta)}\sqrt{1-4\eta}\left({\cal Q}_1 \chi_1^2-{\cal Q}_2 \chi_2^2\right)\right.\nonumber\\
&&\qquad \left.\left(\frac{443}{4(1-2\eta)}-\frac{9355+1008\eta}{14}\right)\sigma_{\rm QM}\right].
\end{eqnarray}

The equation-of-state imprint in the gravitational-wave signals that has received the most attention in recent years is due to
tidal effects. As the quadrupole-monopole term, these are Newtonian effects but they scale with the orbital velocity $v$ as $O(v/c)^5$
and therefore become important later in the inspiral. The neutron star deforms in response to the companion's
nonuniform gravitational potential across its mass distribution. Similar to the rotationally-induced buldge, the tidal bulges distort the object's exterior
gravitational field, which in turn affects the orbital motion and gravitational wave emission. The dominant effect is
characterized by a tidally induced quadrupole scalar of the form
\begin{equation}
{\rm Q}_{\rm tidal}\sim -\Lambda(m, {\rm EOS}) m^5{\cal E},
\end{equation}
where ${\cal E}$ is the companion's
tidal field. In Newtonian gravity, ${\cal E}\sim -m_{\rm comp} /r^3$ but this is generalized for relativistic systems to a definition in terms of the Riemann tensor characterizing the spacetime curvature produced by the companion. The coefficient $\Lambda$ is the
dimensionless tidal deformability parameter, which vanishes for black holes, $\Lambda_{\rm BH}=0$.
The adiabatic quadrupolar tidal
effects give the following contribution to the TaylorF2 phasing \cite{Vines:2011ud}:
\begin{eqnarray}
\lambda_{10,0}&=&-\frac{117}{256\eta^2}\, \tilde \Lambda\,(\pi \mathcal{M} f_0)^{5/3},\\
 \lambda_{12,0}&=& \frac{5(\pi \mathcal{M} f_0)^{7/3}}{512\eta^{12/5}}\left[\frac{3957}{91}\sqrt{1-4\eta}\, \delta \tilde\Lambda-\frac{1869}{16}\, \tilde \Lambda \right].
\end{eqnarray}
Here, $\tilde \Lambda$ and $\delta \tilde \Lambda$ are combinations of the individual tidal parameters given by
\begin{eqnarray}
\tilde \Lambda&=&\frac{16}{13}\sum_{i=1}^2\Lambda_i\frac{m_i^4}{M^4}\left(12-11\frac{m_i}{M}\right)\\
\delta \tilde \Lambda&=& \left(\frac{1690}{1319}\eta-\frac{4843}{1319}\right)\left(\frac{m_1^4}{M^4}\Lambda_1-\frac{m_2^4}{M^4}\Lambda_2\right)\nonumber\\
&&+\frac{6162}{1319}\sqrt{1-4\eta}\left(\frac{m_1^4}{M^4}\Lambda_1+\frac{m_2^4}{M^4}\Lambda_2\right).
\end{eqnarray}

We note that for neutron stars, it was shown that the dimensionless parameters ${\cal Q}$ and $\Lambda$,
characterizing the spin- and tidally-induced quadrupolar deformations of the neutron star's exterior spacetime,
encode similar equation-of-state information and can be
related in an approximately equation-of-state independent way~\cite{Yagi:2013bca,Yagi:2013awa}.

\subsection{Characteristic parameters}

\begin{figure}
  \centering
  \begin{minipage}[t]{1.0\linewidth}
    \includegraphics[width=0.49\linewidth]{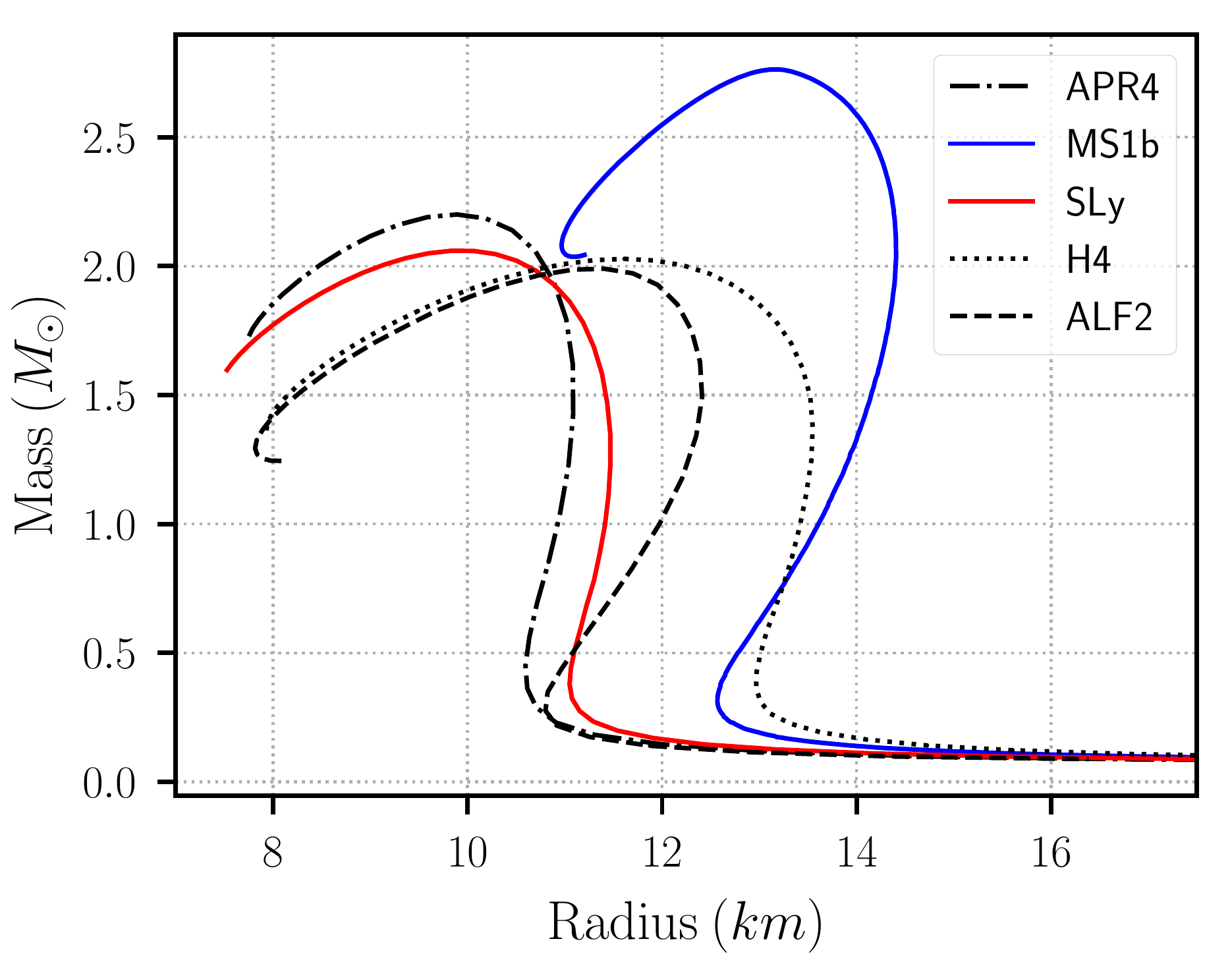}
    \hspace{0.02\linewidth}
    \includegraphics[width=0.49\linewidth]{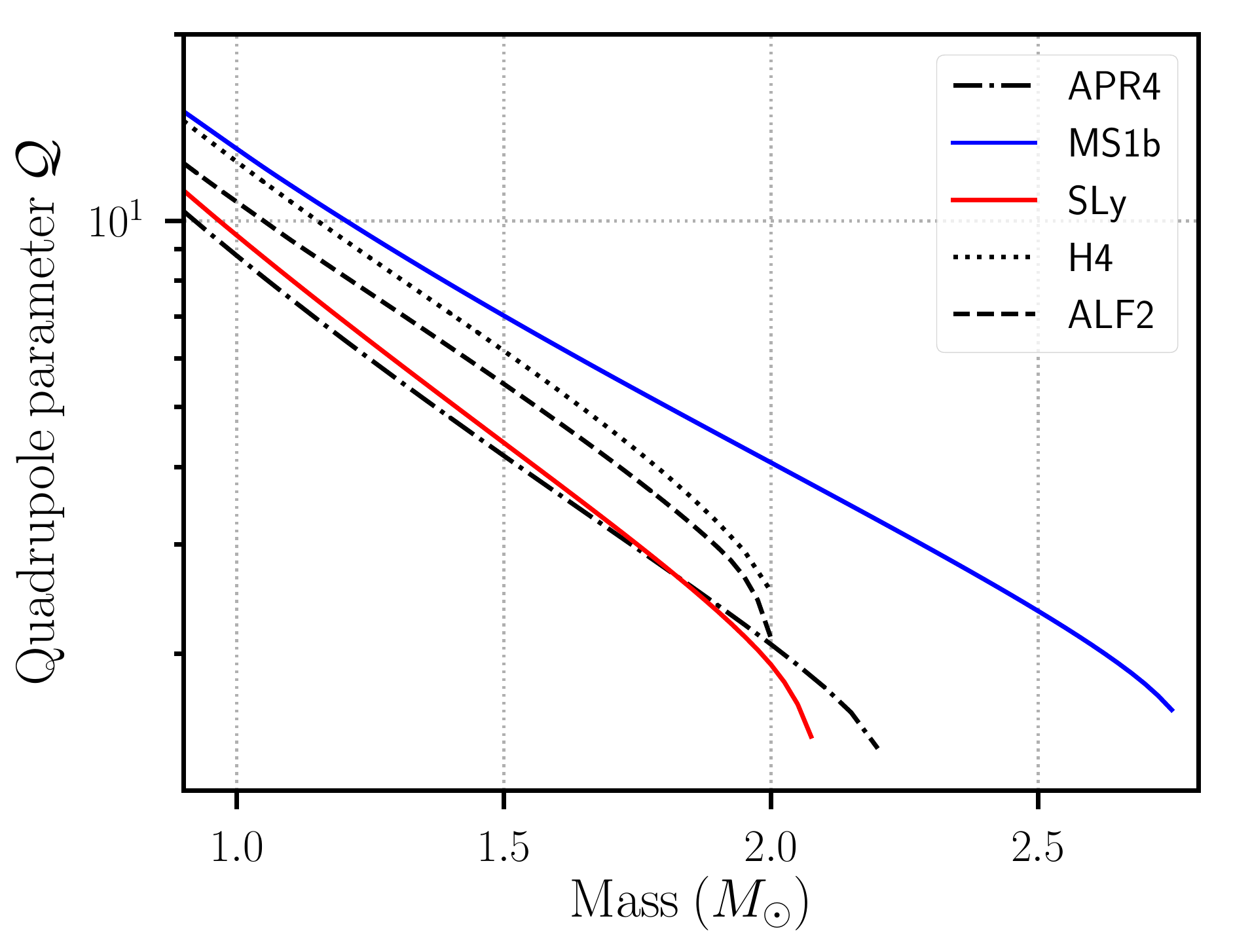}
  \end{minipage}
  \begin{minipage}[t]{0.49\linewidth}
    \includegraphics[width=\linewidth]{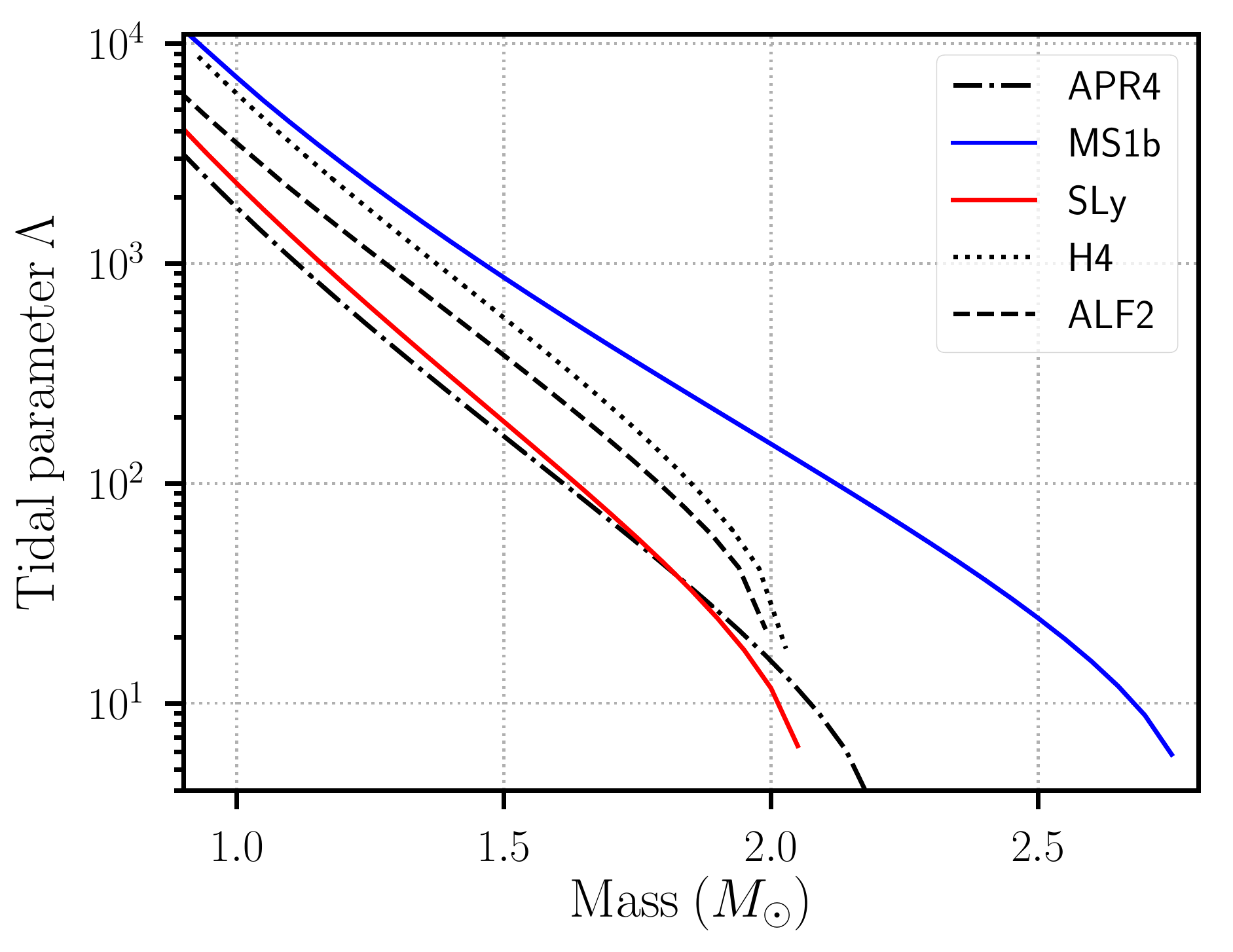}
  \end{minipage}
\caption{\label{fig:parameters}\emph{Parameters characterizing the properties of neutron stars for different equation-of-state models.}
The solid, colored curves are the results from the two fiducial models SLy (red) and MS1b (blue).
The black curves show alternative equation-of-state models. Top left: mass-radius relation, top right: dimensionless
spin-quadrupole parameter, bottom: dimensionless tidal deformability. }
\end{figure}

To illustrate the features of the parameters ${\cal Q}$ and $\Lambda$ we consider representative
examples of proposed equation-of-state models. The equation-of-state of neutron star matter has long remained a scientific challenge,
despite theoretical advances and improved constraints from nuclear experiments and astrophysics.
While most models largely agree up to densities around nuclear density, the large extrapolations
required to apply known nuclear physics to the extreme conditions in neutron star interiors result in a wide range of possible
equations-of-state. Among the numerous
candidate equation-of-state models we consider two cases: one where matter is more compressible and the neutron star
is compact (SLy,~\cite{Chabanat:1997un,Chabanat:1997qh}) and a model where matter is stiff and the neutron star thus
has a large radius for a given mass (MS1b~\cite{Mueller:1996pm}).
We note that the latter model is already disfavored by the preliminary data results
from analysis of the first binary neutron star observation~\cite{TheLIGOScientific:2017qsa}, but is not yet confidently
ruled out, and for our purposes will serve as an upper bound on the size of the matter effects. The two
different equations of state lead to different global parameters of the neutron star, as
shown in figure~\ref{fig:parameters} for the radius, rotational quadrupole parameter, and tidal
deformability, as a function of the neutron star's mass. For comparison, we also included a few
other equation-of-state models that assume a different composition such as hyperons or quark matter at high density,
different values of parameters, or use different methods of calculation. As seen from the plot, the examples SLy and MS1b bracket a range of plausible equations-of-state. 

\begin{figure}
\centering
\includegraphics[width=0.5\textwidth]{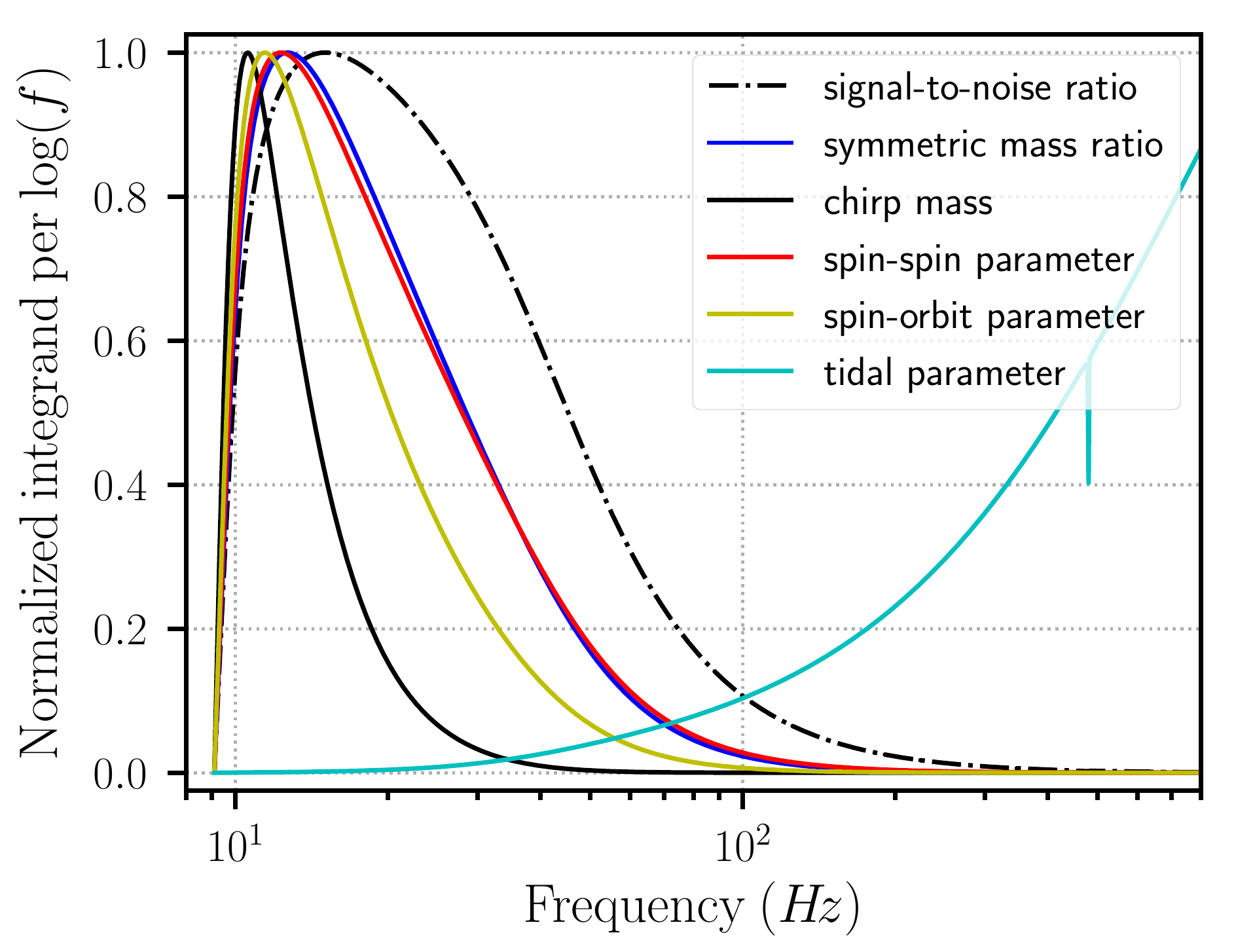}
\caption{\label{fig:info}\emph{Illustration of where in frequency the information about intrinsic binary parameters
predominantly comes from}. The quantity shown on the $y-$axis is a normalized quantity characterizing
the accumulation of information about the binary parameters $\xi_i$ per logarithmic frequency interval.
Specifically, the y-axis is $|(\partial \tilde{h}/\partial {\xi_i})|^2/(f\, S_n)$ for $S_n$ the zero-detuned
high power configuration of Advanced LIGO and each curve normalized to its maximum value.}
\end{figure}

An illustration of where the effects discussed above become important for signals observable by
Advanced LIGO is shown in figure~\ref{fig:info}. The plot depicts 
the normalized accumulation of information about parameters per logarithmic frequency interval
versus the gravitational-wave frequency for an equal-mass binary. The quantity shown in the plot is the normalized value of
$|(\partial \tilde{h}/\partial {\xi_i})|^2/(f\, S_n)$, where $\xi_i=({\cal M},\eta, \beta, \sigma, \tilde \Lambda)$
are the intrinsic source parameters and $S_n$ is the noise power spectral density, for which  Advanced LIGO's zero-detuned high-power configuration~\cite{noisecurve} was used. 
The significance of this quantity for measuring the
parameters will be explained in detail in the subsequent sections. It is the integrand for
the diagonal elements in the Fisher information matrix, up to the factor of $1/f$ which converts
to a logarithmic frequency interval. Each curve is normalized to its individual maximum value,
except for the tidal parameter which is normalized by its value
at a reference frequency of $1$kHz. 
We observe that for the mass-ratio and spin parameters $(\eta, \beta, \sigma)$ the major contribution to the information comes from similar frequency
ranges, while information about the tidal parameter accumulates at much higher frequencies.
This is an important feature that we will return to when discussing our results. Note that in the
Post-Newtonian waveform the symmetric mass ratio $\eta$ first enters at a lower order than the
spin parameters so that one might expect the information about $\eta$ to be concentrated at lower
frequencies. But because $\eta$ also enters at all higher post-Newtonian orders, the distribution in
the plot is shifted to higher frequencies than the spin parameters, for which only the leading order
effect was included to generate this plot. 

%Note also that only individually normalized distributions are shown to enable plotting on the same scale, so one can't infer anything about the absolute info about the parameters. 

%For Zero-Detuned-High-Power configuration. Normalized integrands means the integrands for $(h_{,\xi_i}|h_{,\xi_i})$ per logarithmic frequency interval (see below for the definition of the inner product, commas denote partial derivs and $\zeta_i$ are the signal parameters). The normalization is to the peak values for $({\cal M}, \eta,\beta,\sigma)$ but $\tilde \Lambda$ doesn't have a peak just continues to grow so it's normalized to its value at 1000 Hz. 

Additional information about the equation-of-state can come from the frequency at which the merger or 
tidal disruption occurs, where the latter is mainly relevant for mixed neutron star-black hole binaries. The TaylorF2 waveforms considered here describe only the inspiral portion of the signal and 
are usually terminated at the frequency corresponding to the inner-most stable circular orbit (ISCO) of a
nonspinning system of a test particle orbiting a Schwarzschild black hole with the given masses~\cite{Allen:2005fk}.
We will also test the impact of not using the ``ISCO"-criterion but instead a fit from numerical relativity
simulations for the merger frequency to terminate the inspiral signal~\cite{Bernuzzi:2015rla}.

\section{A brief recap of binary-neutron star data analysis techniques}
\label{sec:searchintro}

In this section we provide a brief recap of a number of the data analysis techniques that we will use in later sections
in this work. For a more complete introduction to these topics we refer the reader to~\cite{Maggiore:1900zz,Creighton:2011zz}.
Consider a stretch of data $s(t)$, recorded by a gravitational-wave observatory. This data is assumed to consist
of colored, Gaussian noise $n(t)$ with the possible presence of a gravitational-wave signal $h(t)$. The noise
is described by the one-sided noise power-spectral density $S_n(f)$, defined by
\begin{equation}
 \frac{1}{2} \delta(f-f') S_n(f) = \mathbb{E}[ \tilde{n}(f) \tilde{n}^{*}(f')], 
\end{equation}
where $\mathbb{E}[\cdot]$ denotes the expectation value over
independent noise realizations. We denote these assumptions of the noise properties with $I$.
When evaluating the likelihood of a signal $h(t)$ being present in the detector data,
one can determine the probability of obtaining the given data realization if no signal is present, $P(s|n,I)$, compared to
the probability of obtaining the same data if a signal is present, $P(s|h+n,I)$. These probabilities can be calculated
according to~\cite{Maggiore:1900zz,Creighton:2011zz}
\begin{equation}
 P(s|h+n,I) \propto e^{-  \left\langle s-h | s-h\right\rangle /2 },
\end{equation}
which reduces to $P(s|n)$ in the case that $h=0$. Here $\left\langle a | b \right\rangle$ defines a
noise-weighted inner product according to
\begin{equation}
\left\langle a|b \right\rangle = 4 \, \mathrm{Re} \int^{\infty}_0\dfrac{\tilde{a}(f)\tilde{b}^*(f)}{S_n(f)} df,
\end{equation}
where $\tilde{a}$ represents the Fourier transform of $a$.
Then the relative probability of the two hypotheses is given by
\begin{equation}
 \frac{P(s|h+n,I)}{P(s|n,I)} = L= e^{ \left\langle s | h\right\rangle - 0.5 \left\langle h | h\right\rangle }.
\end{equation}
This can be maximized over the unknown amplitude of the signal, $A$, to give the matched-filter signal-to-noise
ratio that is routinely used in searches for compact binary mergers~\cite{Allen:2005fk,Babak:2012zx}
\begin{equation}
  2 \log L_{\max A} = \rho^2 = \frac{\left\langle s | h \right\rangle^2}{\left\langle h | h\right\rangle}.
\end{equation}

When attempting to measure the parameters of a known signal in the data, a slightly different question is asked.
We discussed in the previous section how a gravitational-wave signal from a binary neutron star merger depends
on a variety of parameters, which we will collectively denote $\xi_i$. When attempting to measure the parameters
of a known signal one wishes to know what is the probability of the signal having a specified set of parameters,
$P(\xi_i | s,I)$. According to Bayes' Theorem this is given by
\begin{equation}
  P(\xi_i | s,I) = \frac{P(\xi_i | I)}{P(s | I)} P(s | \xi_i, I).
\end{equation}
The term $P(s , I)$ in this application acts only as a normalization factor such that $\int P(\xi_i | s, I) d \xi = 1$.
The quantity $P(\xi_i | I)$ represents the prior belief that some values of the signal parameters $\xi_i$ are expected to be more likely than
others. Finally $P(s | \xi_i, I)$ represents the probability of obtaining a
specific noise realisation $s$ given a specific choice of parameters and our stated assumptions $I$.
If one can evaluate $P(\xi_i | s,I)$ at all possible values of $\xi_i$ one has a direct measurement of the
probability of different parameters. In general the parameter space is too large to allow a direct measurement
and instead techniques to draw samples from the underlying probability distribution are employed~\cite{Veitch:2014wba}.
An alternative, and much quicker, method to compute
the expected bias in the peak of $P(s | \xi_i, I)$ due to underlying noise is to use the Fisher Information
Matrix~\cite{Vallisneri:2007ev}. This
is very quick to evaluate, but one must be careful when using this as it provides an estimation of the matched-filter
between two waveforms $(h(\xi_i) | h(\xi_i + \delta \xi_i))$, which is only valid when $\delta \xi_i$ tends to 0. For this
to be valid for small, but non-negligible values of $\delta \xi_i$, as might be expected when estimating the parameters
of gravitational-wave signals, the underlying parameter space metric must not vary strongly in the parameters used to
evaluate the Fisher Information Matrix~\cite{Vallisneri:2007ev}.
In this work we will instead attempt to measure $P(\xi_i | s,I)$ directly, making some assumptions to
reduce the dimensionality of the parameter space, as we will discuss later in section~\ref{sec:param_bias}.

\section{Waveform mismatch with known masses and component spins}
\label{sec:faithfulness}

In this section we begin our exploration of the effect that equation-of-state dependent terms in
the waveform model can have on gravitational-wave searches with a simple question.
If we assume that all binary neutron star systems have the
same---known---values of component masses and component spins, would it be possible to observe the difference
between binary neutron-star systems with different equations-of-state, or a binary-black hole merger with the same
component masses and spins. If the answer to this question were ``no''
then equation-of-state terms would not be possible to measure with observatories like Advanced LIGO.

To answer this question we note that the matched-filter signal-to-noise ratio between a filter waveform $h$
and a stretch of data $s$ containing noise $n$ and a signal $g$ is a linear sum
\begin{equation}
 \left\langle s | h \right\rangle = \left\langle n | h \right\rangle + \left\langle g | h \right\rangle.
\end{equation}
Assuming the noise is Gaussian and stationary the average value of $\left\langle n | h \right\rangle$ over
multiple noise realizations is 0, so one can consider a ``zero-noise'' realization where $\left\langle n | h \right\rangle=0$
and neglect the noise contribution. Then the normalized matched-filter, or overlap, between the two waveforms
\begin{equation}
\label{eq:overlap}
 \mathcal{O}(g,h) = (\hat{g}|\hat{h}) = \dfrac{(g|h)}{\sqrt{(g|g)(h|h)}},
\end{equation}
gives the fraction of the optimal signal-to-noise ratio that is recovered when searching for a signal $g$
using $h$ as the waveform filter.

Finally, we wish to maximize this quantity over the unknown source orientation and sky location. With
the waveform model we are using, these parameters enter as a combination of an overall phase shift, overall
amplitude shift and overall time-shift~\cite{Buonanno:2009zt,Babak:2012zx}.
Therefore we define the ``match'' as the overlap maximized over
a phase shift and a time-shift, which is easily computed as described in~\cite{Allen:2005fk,Babak:2012zx}
\begin{equation}
\mathcal{M}(h_1,h_2) = \underset{\phi_c,t_c}{\max}(\hat{h}_1|\hat{h}_2(\phi_c,t_c)).
\end{equation}
The value of this match at which signals would be distinguishable depends upon the
signal-to-noise ratio of the signal, as well as the geometry of the parameter space being considered and
any strong priors being used. A simple rule-of-thumb, described in~\cite{Lindblom:2008cm} argues that signals
can be distinguished if the signal-to-noise ratio squared is reduced by an absolute value of 1 when searching
for $h_1$ using $h_2$ as a filter, compared to the optimal signal-to-noise ratio where $h_1$ is used as the filter. For
a signal-to-noise ratio of 15 this corresponds to waveforms being indistinguishable if the match is larger
than 0.9978, or for a signal-to-noise ratio of 25, if the match is larger than 0.9992.

We first use this match to determine whether second generation gravitational wave observatories will be sensitive
to variations in the value of the spin quadrupole moment scalar, $\mathcal{Q}$. This is done by
generating two waveforms where the only difference is in the value of $\mathcal{Q}$ and computing the match between them.
By repeating this procedure over a range of masses and spins we can evaluate where in the parameter space it might be possible
to distinguish differences in waveforms due to variations in the spin quadrupole moment scalar. Here
we set the tidal deformability parameter, $\Lambda$, of both bodies to 0 and for all waveforms use a termination frequency
corresponding to the black hole ``ISCO"-criterion as described in section~\ref{sec:waveforms}.

\begin{figure}[t!bp]
  \centering
  \begin{minipage}[t]{1.0\linewidth}
    \includegraphics[width=0.49\linewidth]{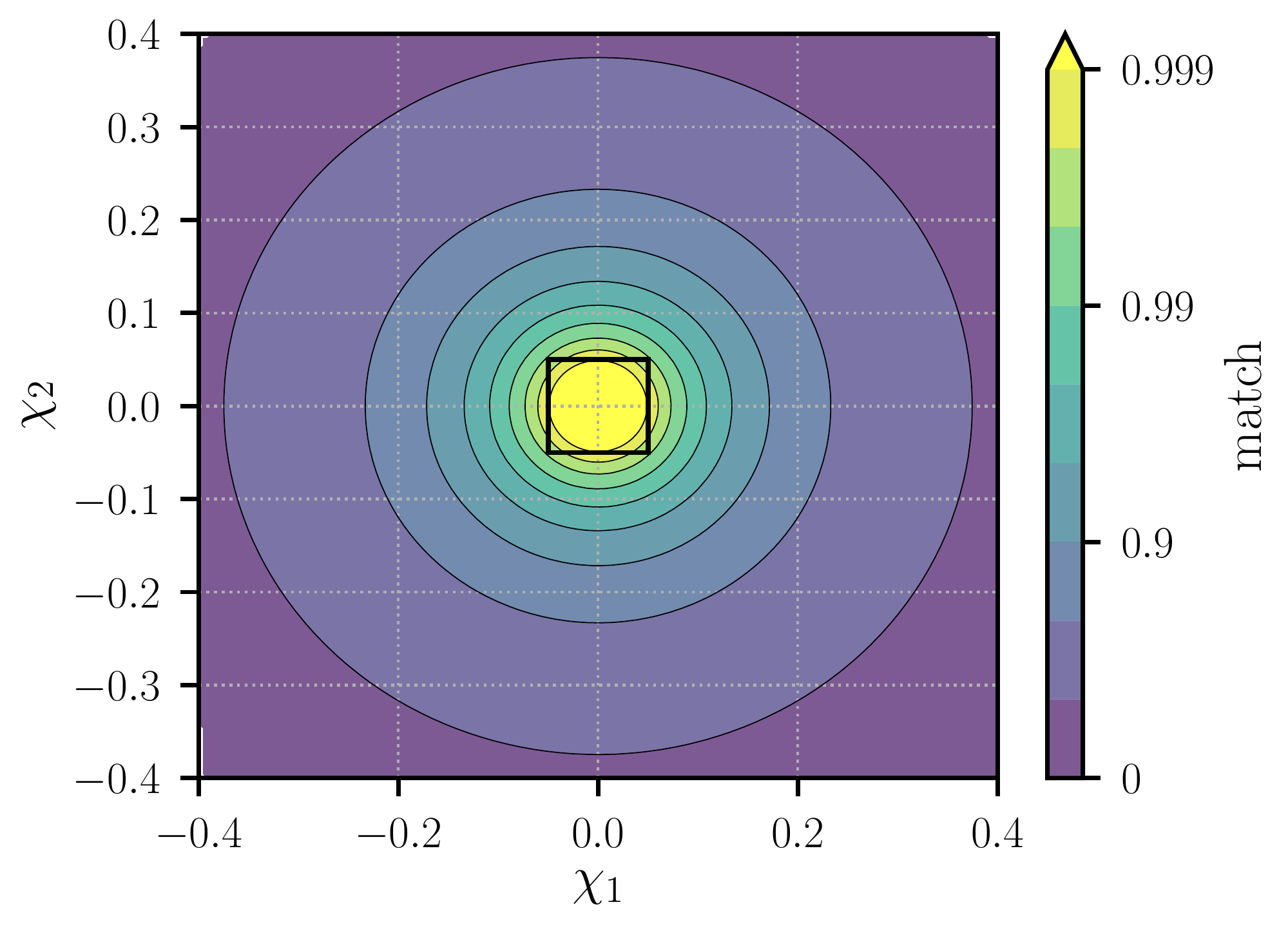}
    \hspace{0.02\linewidth}
    \includegraphics[width=0.49\linewidth]{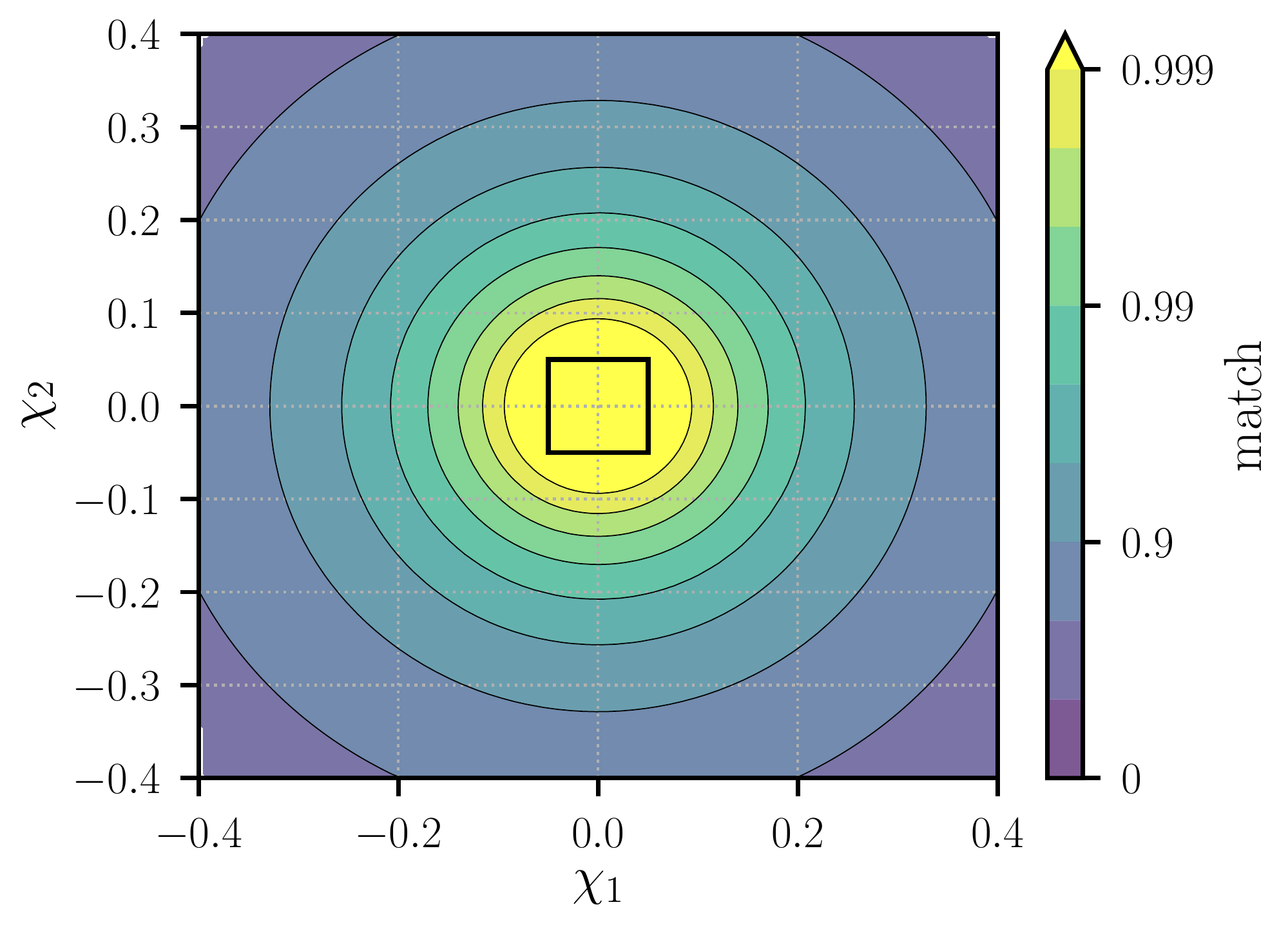}
  \end{minipage}
\caption{\label{fig:faithfulness1}
The match, as a function of the component spins, between waveforms where both neutron stars are modelled
with a ${\cal Q}$ parameter of 1 and a ${\cal Q}$ parameter of 12 (left) or 4 (right). Here both bodies have
a component mass of $1.35 M_{\odot}$ and do not include any effect due to tidal deformation. Matches are
computed using the Advanced LIGO zero-detuned, high-power noise sensitivity curve.}
\end{figure}

In figure~\ref{fig:faithfulness1}, we show the match, as a function of the component spins,
between waveforms with ${\cal Q}=1$ and waveforms modelled with either ${\cal Q} = 4$ or ${\cal Q} = 12$. 
These fiducial values are chosen as examples from the range exhibited in
figure~\ref{fig:parameters}, e.g. ${\cal Q}\sim 4$ for the SLy model at $m\sim 1.6M_\odot$,
and ${\cal Q}\sim 12$ for the MS1b model at $m\sim 1.1 M_\odot$. The component masses here are chosen
to be $1.35 M_{\odot}$ for both bodies although we note that the plots look qualitatively similar when using different
component masses and the same values of ${\cal Q}$. From these plots we can see that the
effect of the neutron-star self-spin deformation will have a negligible effect on the emitted gravitational-wave signal if
the dimensionless spins of both bodies are less than $\chi=0.05$, as would be expected for non-recycled neutron stars.
However, if we consider neutron-star systems with spins as large as $\chi=0.4$, as might be possible with recycled
neutron stars, then the self-spin deformation causes very large mismatch between waveforms.
Exotic compact objects can have much larger ${\cal Q}$ than neutron stars (see e.g. Ref.~\cite{Ryan:1996nk} for the case of boson stars),
which would give a much more noticeable effect.

\begin{figure}[t!bp]
  \centering
  \begin{minipage}[t]{1.0\linewidth}
    \includegraphics[width=0.49\linewidth]{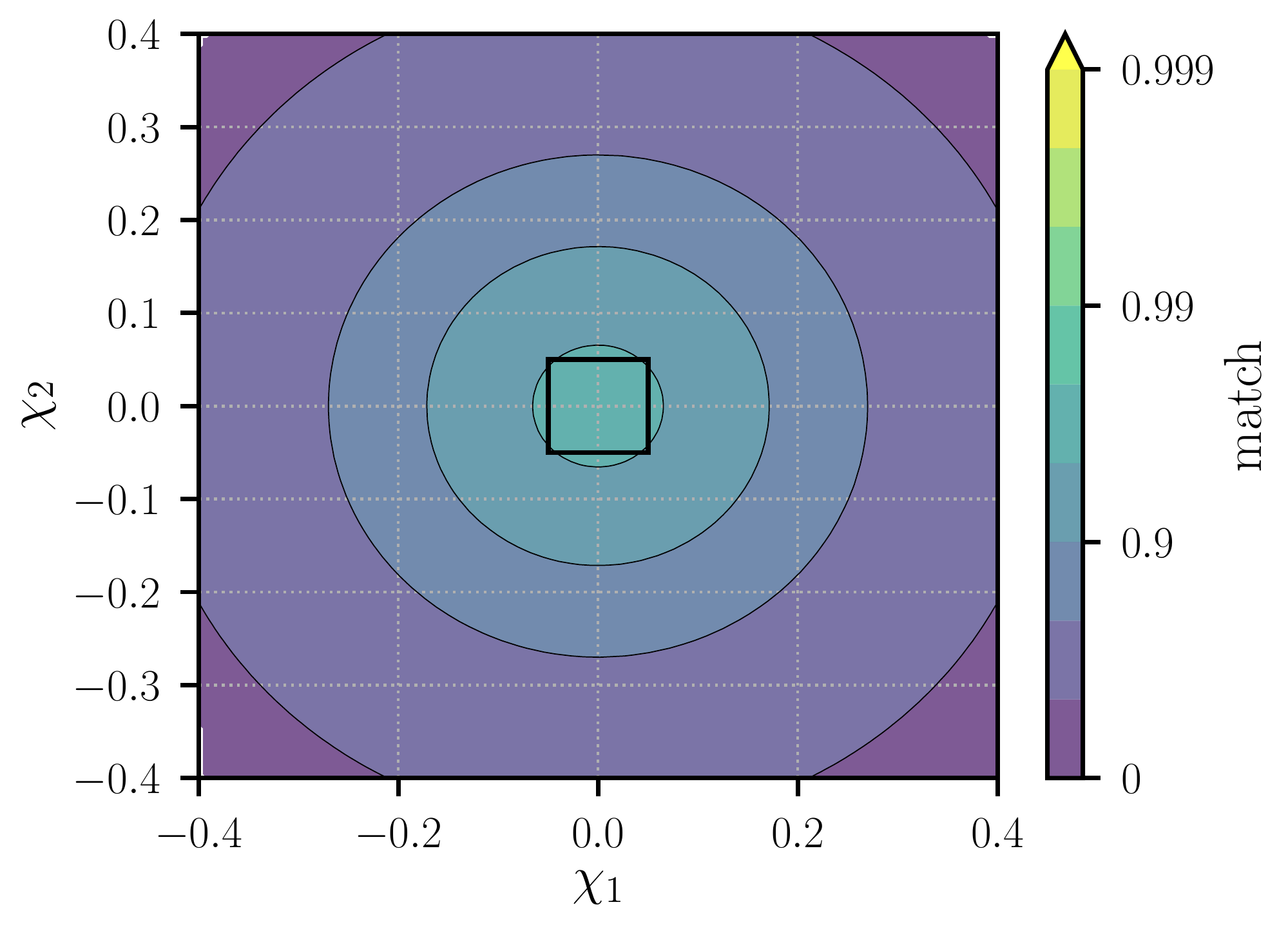}
    \hspace{0.02\linewidth}
    \includegraphics[width=0.49\linewidth]{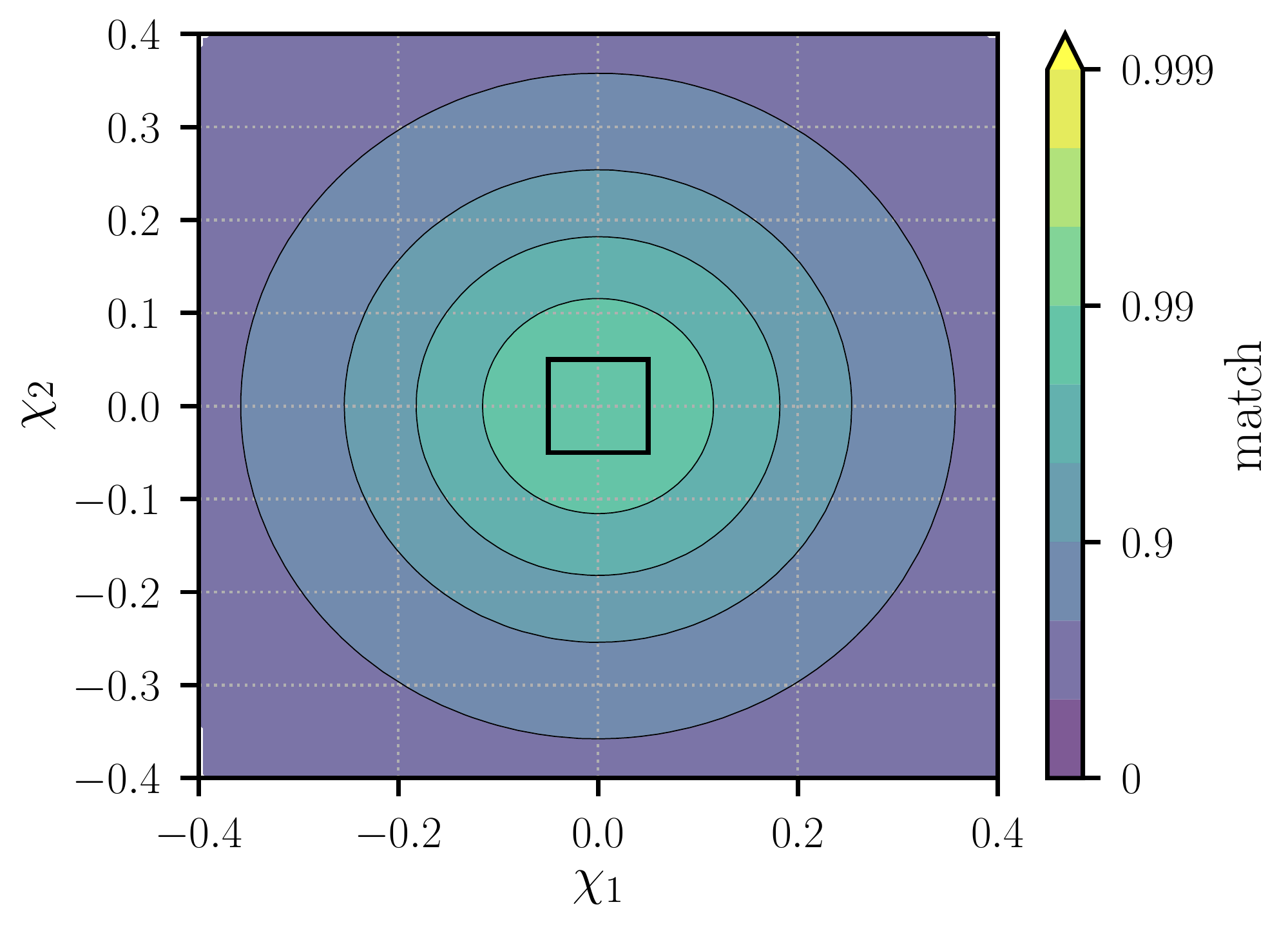}
  \end{minipage}
  \centering
  \begin{minipage}[t]{1.0\linewidth}
    \includegraphics[width=0.49\linewidth]{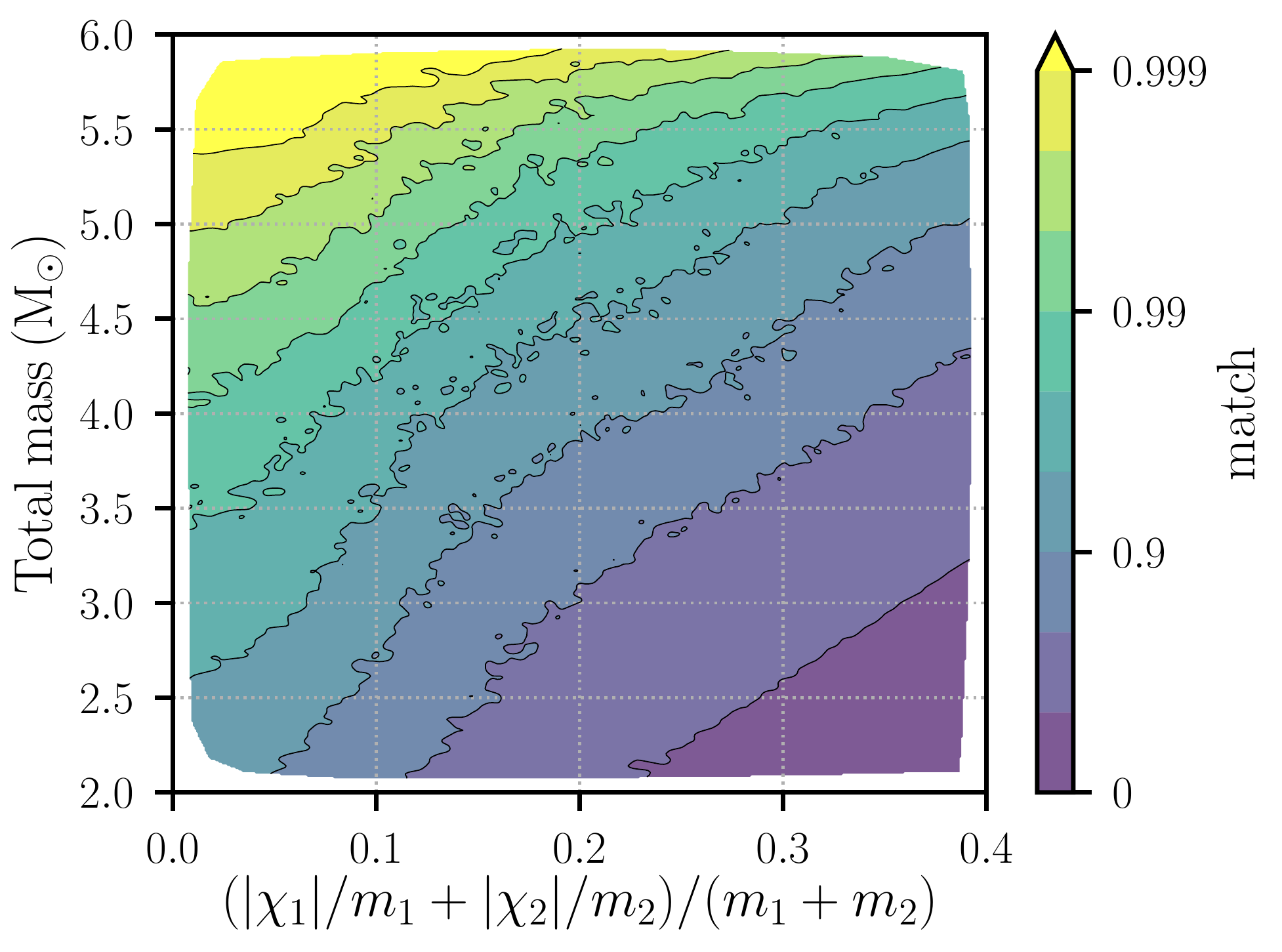}
    \hspace{0.02\linewidth}
    \includegraphics[width=0.49\linewidth]{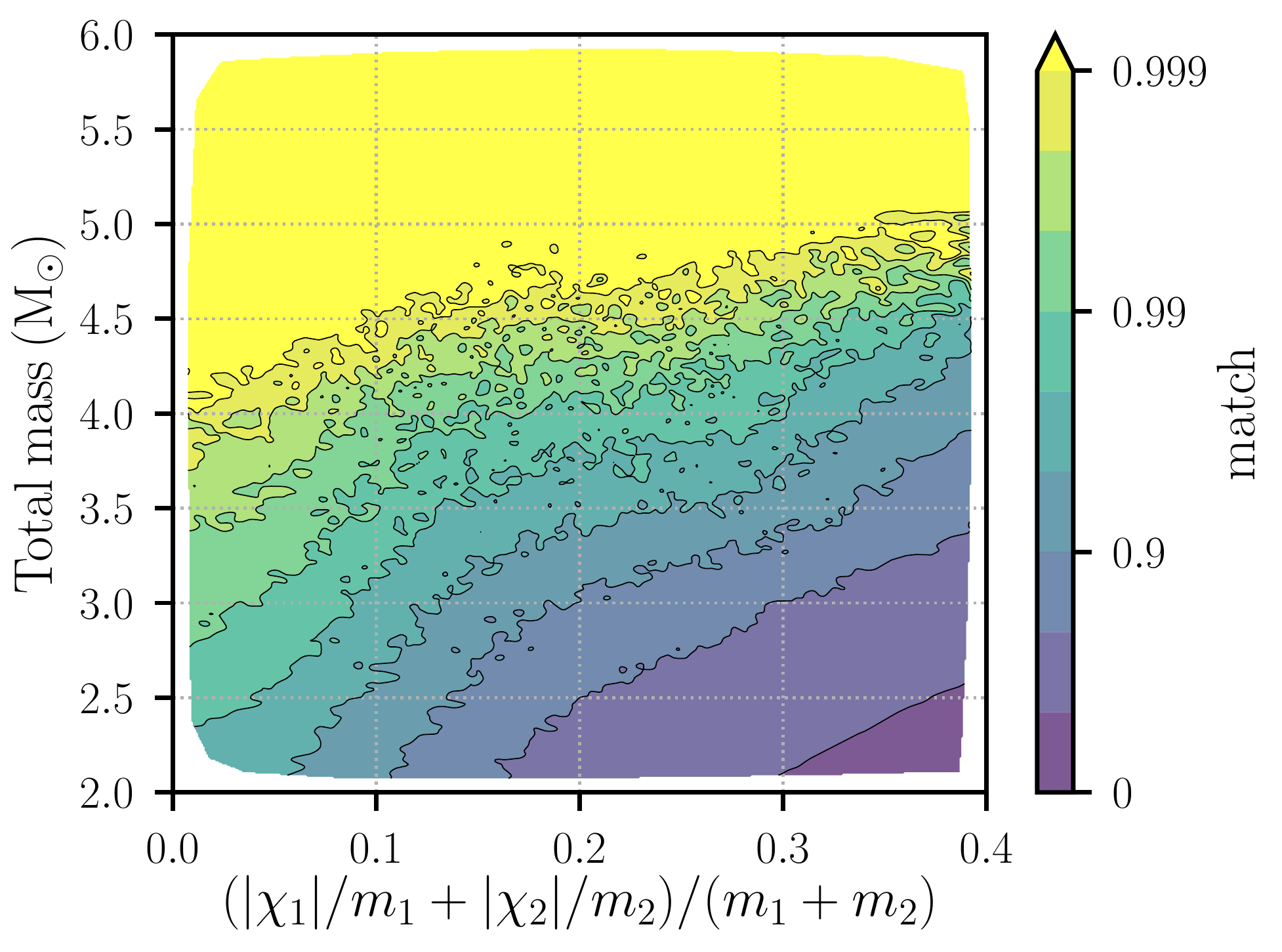}
  \end{minipage}
\caption{\label{fig:faithfulness2}
The match between waveforms modelled as two black holes and waveforms, with the same component
masses and spins, modelled as two neutron stars using either the MS1b or SLy equations of state.
Top: Match shown as a function of the two component spins for binaries where both component
masses are 1.35 using the MS1b (left) or SLy (right) equation of state. Bottom: Match shown
as a function of the total mass and a mass-weighted spin contribution using the MS1b(left)
or SLy (right) equation of state. The bottom plots are 2-dimensional smoothed projections of the match
calculated as a function of the two masses and two spins. As there is some variation of the match in the dimensions
projected away, some noisiness is present in this plot.
In all cases the Advanced LIGO zero-detuned, high-power noise
sensitivity curve is used.}
\end{figure}

In figure~\ref{fig:faithfulness2} we show the match, as a function of the component spins, between waveforms modelled
assuming that both bodies are binary black holes and waveforms modelled assuming both bodies are neutron stars described
by a given equation of state. Here we use both the MS1b and SLy equations of state, described in section \ref{sec:waveforms}.
These waveforms include not only the effect of the components' spin-quadrupole term, but
also tidal terms and a difference in the termination frequency.
In the top panels of figure~\ref{fig:faithfulness2}, as in figure~\ref{fig:faithfulness1},
we choose $m_1 = m_2 = 1.35 M_{\odot}$ for all cases.
Here we see that at large values of the component spins the mismatch is dominated by the value of ${\cal Q}$, which takes
a value of 8.39 for both bodies for these masses with the MS1b equation of state, and 5.54 with the SLy equation of state.
However, when the component spins tend to 0 the mismatch does not approach 0. In this case the mismatch is dominated
by the presence of the tidal deformation term characterized by $\Lambda$ which for these masses takes a value of 1510 for both bodies
with the MS1b equation-of-state and 382 with the SLy equation of state. For systems with zero spins and these masses the
match is 0.960 for the MS1b equation of state and 0.990 for the SLy equation of state.
The difference in termination
frequency reduces the match by only 0.0006 in the case of MS1b, which is much smaller than the contribution from the tidal deformation.
However, the $\Lambda$ and ${\cal Q}$ terms are strongly mass dependent, as seen from figure~\ref{fig:parameters}, and will
be larger at lower masses, and smaller at higher masses.
In the bottom panels of figure~\ref{fig:faithfulness2}
we have computed matches for signals chosen with component masses uniformly distributed between 1 and 3 solar masses, and
with component dimensionless spins distributed uniformly between -0.4 and 0.4. The results from this complete
4-dimensional parameter space is then plotted as a smoothed projection into the two dimensional parameter space of total mass
and a mass-weighted spin term. There is some small variation of the match in the two dimensions projected away, which
causes some noisiness in the smoothed plot, but the general trend
is clear. Here we can clearly see that for increasing values of total mass and increasing values
of the component spins the equation-of-state dependent terms become increasingly important. Points close to 0 on the x-axis
will have very little contribution from the spin-quadrupole terms, and the mismatch here is mainly due to the tidal deformation.
The decreasing mismatch as the component spins increase is due to the increasingly important contribution of the spin-quadrupole terms.

These results demonstrate that the commonly-considered tidal terms are the dominant effect arising from the neutron-star
matter when the component spins are small. However, if the spins are large, as would be expected for recycled
neutron stars, the spin-quadrupole terms are the dominant equation-of-state related term and cause significant mismatches
between otherwise identical waveforms. Therefore the spin-quadrupole terms must not be neglected
when considering neutron-star systems with large spins.

\section{Waveform mismatch with unknown masses and component spins}
\label{sec:effectualness}

From our results for the observability of effects in the gravitational-wave signals discussed above and displayed in
figures~\ref{fig:faithfulness1} and ~\ref{fig:faithfulness2},
we already drew three conclusions: equation-of-state dependent effects
can have a significant affect on neutron-star waveforms, for slowly-spinning binaries the tidal term is the
dominant equation-of-state effect, and for rapidly spinning neutron stars the spin quadrupole term will have a
much larger effect than tides. 
However, these results 
do \emph{not} allow us to make the conclusion that the presence of the equation-of-state dependent terms will enable us to measure
the equation-of-state in an observation of a binary-neutron-star system. The reason for this is that in general
the masses and spins of a binary neutron star system are not known a-priori, and so these must also be measured in
combination with the equation-of-state related terms. 

In
this section we assume that the component masses and spins are
not known and ask if there would be a loss in the optimal
signal-to-noise ratio if searching for binary neutron star systems
using waveforms with incorrect equation-of-state parameters when the signal-to-noise ratio is maximized over
the unknown component masses and spins. This measure is
important for two reasons, to quantify if there is a significant reduction
in the obtained signal-to-noise ratio if using the wrong
equation-of-state parameters, and if searching for neutron star binary
mergers with black hole templates will lead to a reduction in the overall number
of binary neutron star mergers that would be observed.
If it is not possible to find a waveform with a match
close to unity when using an incorrect set of equation-of-state
parameters and after maximizing over the unknown component
masses and spins it would imply that we would be able
to distinguish between the two equations-of state. In this section
our focus will be on the question of how using wrong
equation-of-state parameters could cause a reduction in the
number of observations being made. The impact of neglecting
equation-of-state dependent terms on detection rate has
been studied in previous works in the context of non-spinning neutron-star systems~\cite{Cullen:2017oaz}. We address here, for the first time,
this question in the context of binary neutron-star systems with component spins as large as 0.4, where the spin-quadrupole and
the tidal deformation terms are considered.

To be able to answer this question we need to calculate what fraction of the signal power is lost after maximizing
over the mass and spin parameters. When searching for compact binary mergers a discrete set of waveforms,
$b_i$, is used~\cite{Sathyaprakash:1991mt,Cokelaer:2007kx,Capano:2016dsf}.
The ``fitting factor'' (as first defined in \cite{Apostolatos:1995pj}) is the maximum overlap between the set of waveform filters and
a potential signal waveform $h$
\begin{equation}
\mathrm{FF}(h, b_i) = \max_{i} \mathcal{M}(h, b_i).
\end{equation}
Normally in a search one creates the set of waveforms, $b_i$, to fulfill the criterion that a signal anywhere in
the parameter space being covered would have a fitting factor of at least 0.97. However, if signals are not
contained within the parameter space being considered, for example if they contain equation-of-state terms not
included in $b_i$, the obtained fitting factor can be lower than the expected minimum. A standard practice
to evaluate this~(\cite{Apostolatos:1995pj,Capano:2013raa,Privitera:2013xza} for example) is to compute 
the fitting factor for a population of signals and plot the distribution. However,
this can sometimes be misleading as often the signals with the lowest values of fitting factors are also
ones whose observable gravitational-wave strain is smallest. Therefore we also define,
following~\cite{Buonanno:2002fy,Harry:2013tca},
the ``signal recovery fraction'' between a population of signals $h_j$ and a discrete set of filter waveforms $b_i$
\begin{equation}
\mathrm{srf}(h_j, b_i) = \frac{ \sum_j \mathrm{FF}^3(h_j, b_i) \sigma(h_j)^3}{\sum_j \sigma(h_j)^3},
\end{equation}
where $\sigma(h_i) = \sqrt{ \left\langle h_i | h_i \right\rangle }$  is proportional to the observable signal power.

In this way the signal recovery fraction gives the fraction of signals, described by the population $h_j$,
that would be recovered above an arbitrary signal-to-noise ratio threshold using the given set of filter waveforms
$b_i$, compared to a theoretical search that includes all possible values of $h_j$ in the set of filter waveforms.
Or, in short, it quantifies what fraction of signals would be missed because of imperfect coverage of the filter waveforms.

\begin{figure}[t!bp]
  \centering
  \begin{minipage}[t]{1.0\linewidth}
    \includegraphics[width=0.49\linewidth]{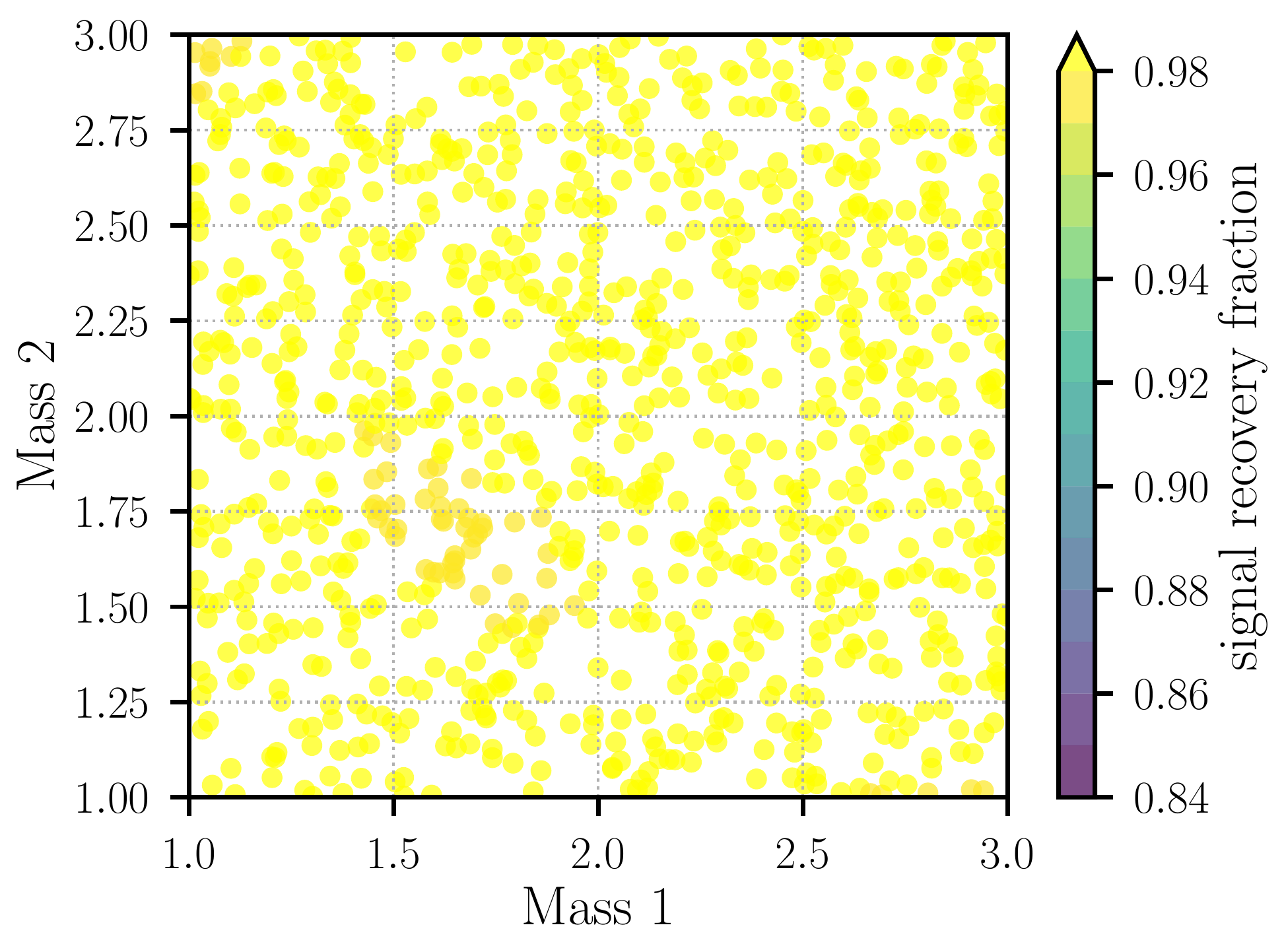}
    \hspace{0.02\linewidth}
    \includegraphics[width=0.49\linewidth]{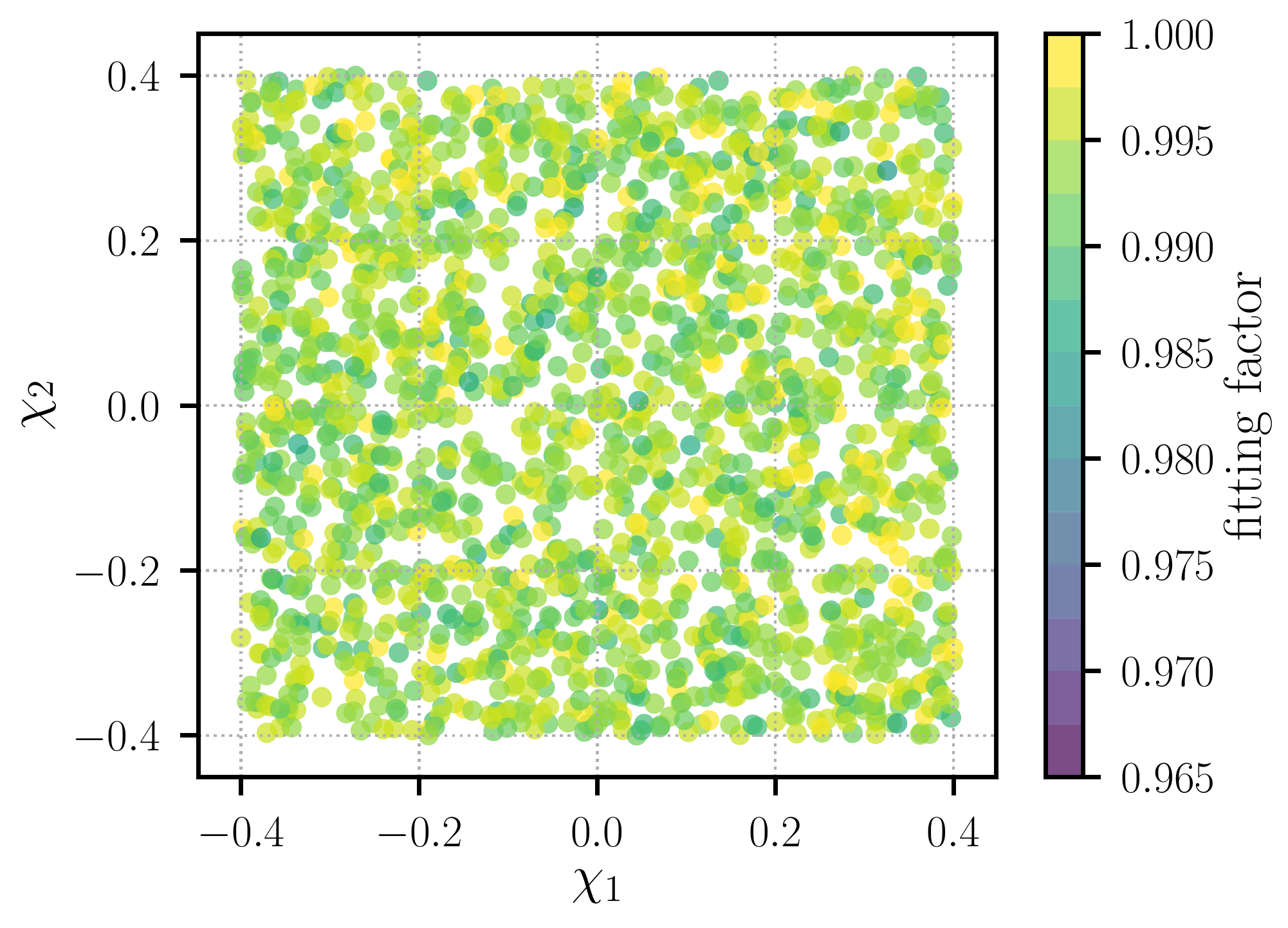}
  \end{minipage}
\caption{\label{fig:effectualness1}
Left: Signal recovery fraction as a function of component masses for a population of
signals modelled as if they were Kerr black holes recovered with a template bank
also containing signals modelled as if both bodies were Kerr black holes. Right: The
fitting factor as a function of the two components' aligned spin between this
template bank and a set of systems with $m_1 = m_2 = 1.35M_{\odot}$.}
\end{figure}

To evaluate fitting factors and signal recovery fractions in this section we first create a set of filter waveforms $b_i$.
Here we create a set of filter waveforms using the methods described in~\cite{Brown:2012qf}. These filter waveforms are constructed
assuming that both bodies are black holes (which during the inspiral means spinning point-masses), with component masses between 1 and 3 $M_{\odot}$ and component dimensionless
spins $\chi \in [-0.4,0.4]$. The set of filter waveforms is constructed such that the maximal loss in signal-to-noise ratio
for any waveform in this parameter space due to the discreteness of the bank is 1\%\footnote{The methods described
in~\cite{Brown:2012qf} for template bank construction make some approximations to the signal model,
which mean that in some parts of the parameter
space the loss in signal-to-noise ratio is a little larger than 1\%, as illustrated in figure~\ref{fig:effectualness1}}.
This number is smaller than the commonly used value of 3\% because, as mentioned above, we choose a smaller value here to emphasize
the effect of the equation-of-state terms.

In figure~\ref{fig:effectualness1} we show the signal recovery fraction
between a population of black holes in this parameter space and our set of filter waveforms.
We choose a distribution of component spins uniform in component spin
magnitude between -0.4 and 0.4 (reminding the reader that we are restricting to only considering aligned-spin systems
in this work). The sky location and orientation of the sources are chosen isotropically, and the signal-recovery fraction
measure already assumes a uniform-in-volume distribution. The signal-recovery fraction is evaluated and plotted as a
function of the two component masses---that is we choose a distinct set of signals and calculate signal recovery fraction
for every point in the component mass space shown in the plots. In all cases in figure~\ref{fig:effectualness1} we see
signal recovery fractions larger than 0.98, which is expected as the waveforms are contained within the parameter space
being considered---we are not yet including equation-of-state effects. We also show the fitting
factor as a function of the two component spins when the component mass of both bodies is 1.4$M_{\odot}$, illustrating
that the fitting factor does vary because of the discrete nature of the bank. This figure provides the benchmark for
the other plots in this section. There, any reduction in signal recovery fraction or fitting factor
from that shown in figure~\ref{fig:effectualness1} is due entirely to the effects of neutron-star physics that can not be
recovered using black-hole waveforms. This would also correspond to the signal loss that is present in current searches
for binary neutron stars in LIGO and Virgo data, where such terms are not currently included.

\begin{figure}[t!bp]
  \centering
  \begin{minipage}[t]{1.0\linewidth}
    \includegraphics[width=0.49\linewidth]{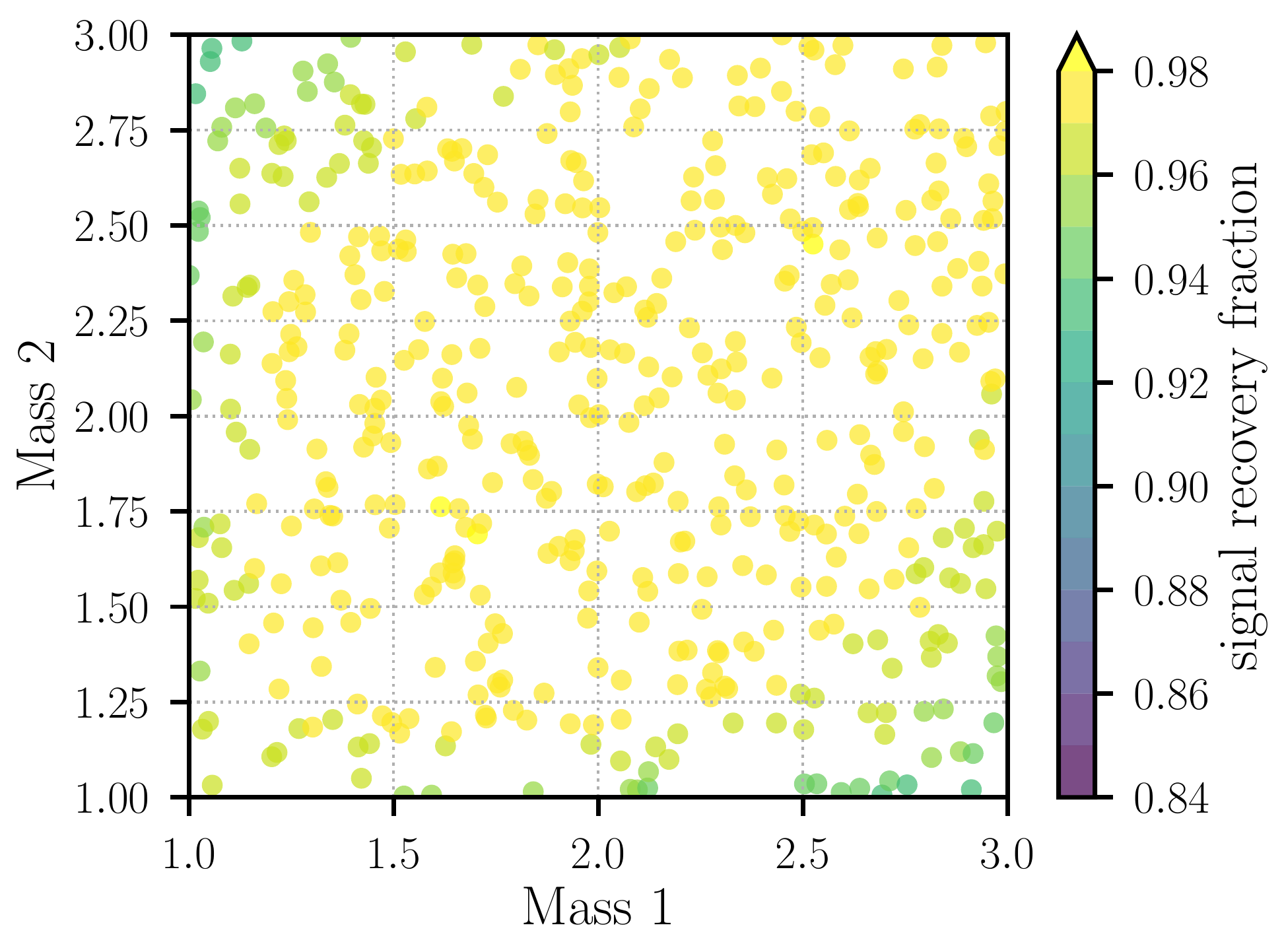}
    \hspace{0.02\linewidth}
    \includegraphics[width=0.49\linewidth]{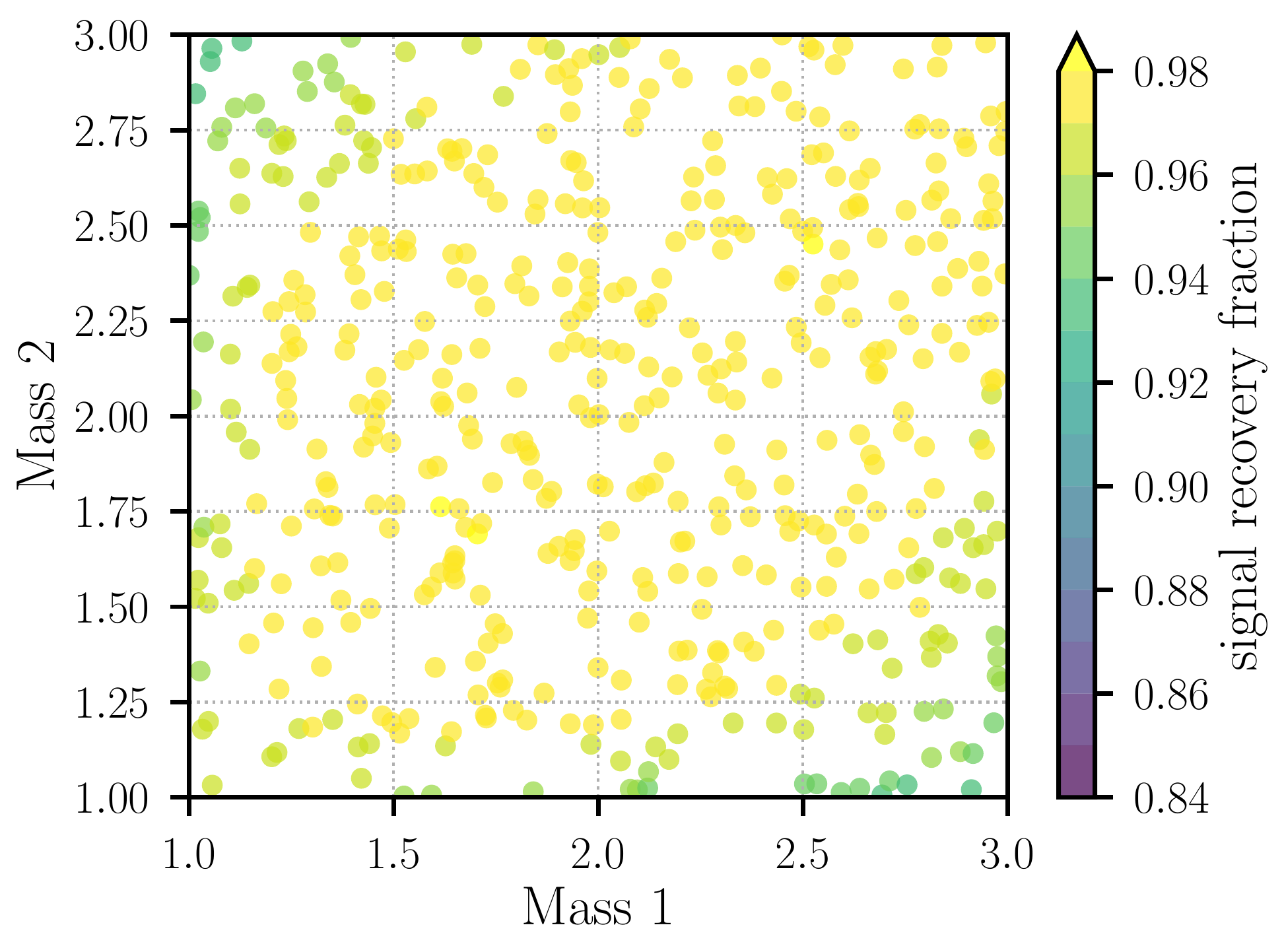}
  \end{minipage}
  \begin{minipage}[t]{0.49\linewidth}
    \includegraphics[width=1.0\linewidth]{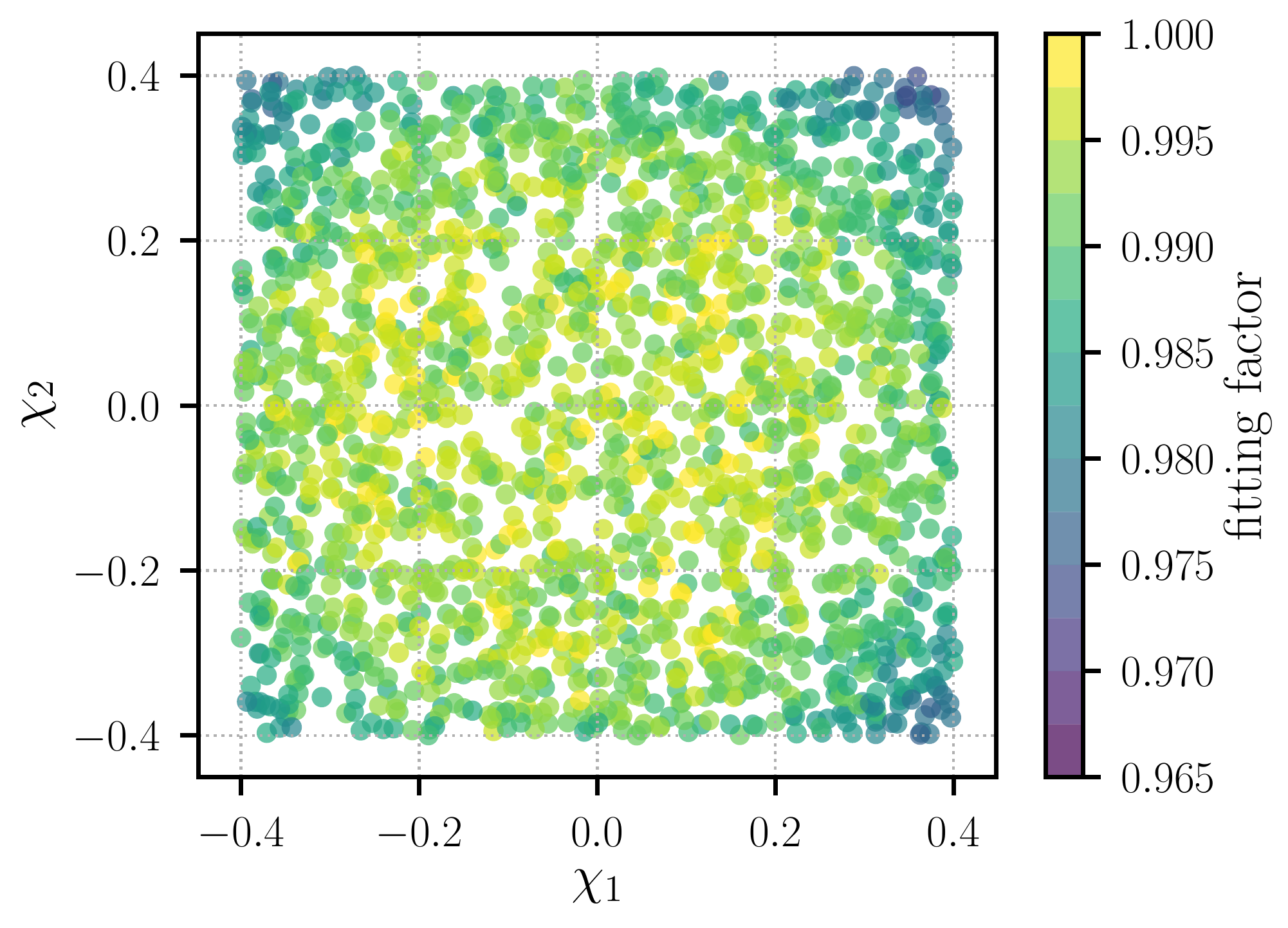}
  \end{minipage}
\caption{\label{fig:effectualness2}
Top left (top right): Signal recovery fraction as a function of component masses for a population of
signals modelled as if they were Kerr black holes, except for the self-spin term,
which is set to 12 (4) for both bodies and recovered with a template bank
containing signals modelled as if both bodies were Kerr black holes. Bottom:
The fitting factor as a fcuntion of the two components' aligned spin between our
Kerr black hole template bank and signals with a self-spin term value of 12, as in the
top panel, but with $m_1 = m_2 = 1.35M_{\odot}$.}
\end{figure}

We first measure the signal-recovery fraction with a set of signals, which deviate from binary-black hole waveforms
only by the inclusion of the quadrupole-monopole term. We perform two sets of simulations, one where ${\cal Q}$ is set to a value
of 4 and one where it is set to a value of 12. In figure~\ref{fig:effectualness2} we measure the signal-recovery fraction using the same
distribution of signals as figure~\ref{fig:effectualness1}. We can see that in most regions in the
component mass parameter space the signal recovery fraction is not noticeably lower than when using low-mass binary-black
hole waveforms. This tells us that in these regions of parameter space the spin-quadrupole term deviations in the waveform
are almost completely degenerate with changes in the component spins and masses, as is also expected based on the discussion
of figure~\ref{fig:info}.
Current Advanced LIGO and Virgo searches would be able to observe signals that produce
waveforms matching the ones used here. At the corners of the parameter space we do notice a significant drop in the signal
recovery fraction. This is because the bank of waveform filters we used does not extend past the component mass and spin
limits quoted above, in this case the signals here would match well with systems outside of the parameter space (ie. with
component masses $< 1$ or $>3 M_\odot$ or component spin magnitudes $>0.4$). We also show the fitting factor as a function of
component spins for systems with both component masses equal to 1.35 $M_{\odot}$ and modeled with ${\cal Q} = 12$.
Here we notice a small reduction in the fitting factor only when the sum of the two component spins is large.

\begin{figure}
  \centering
  \begin{minipage}[t]{1.0\linewidth}
    %\includegraphics[width=0.49\linewidth]{images/kerrtmpls_MS1Bsignals_srf.pdf}
    %\hspace{0.02\linewidth}
    %\includegraphics[width=0.49\linewidth]{images/kerrtmpls_SLYsignals_srf.pdf}
    \includegraphics[width=0.49\linewidth]{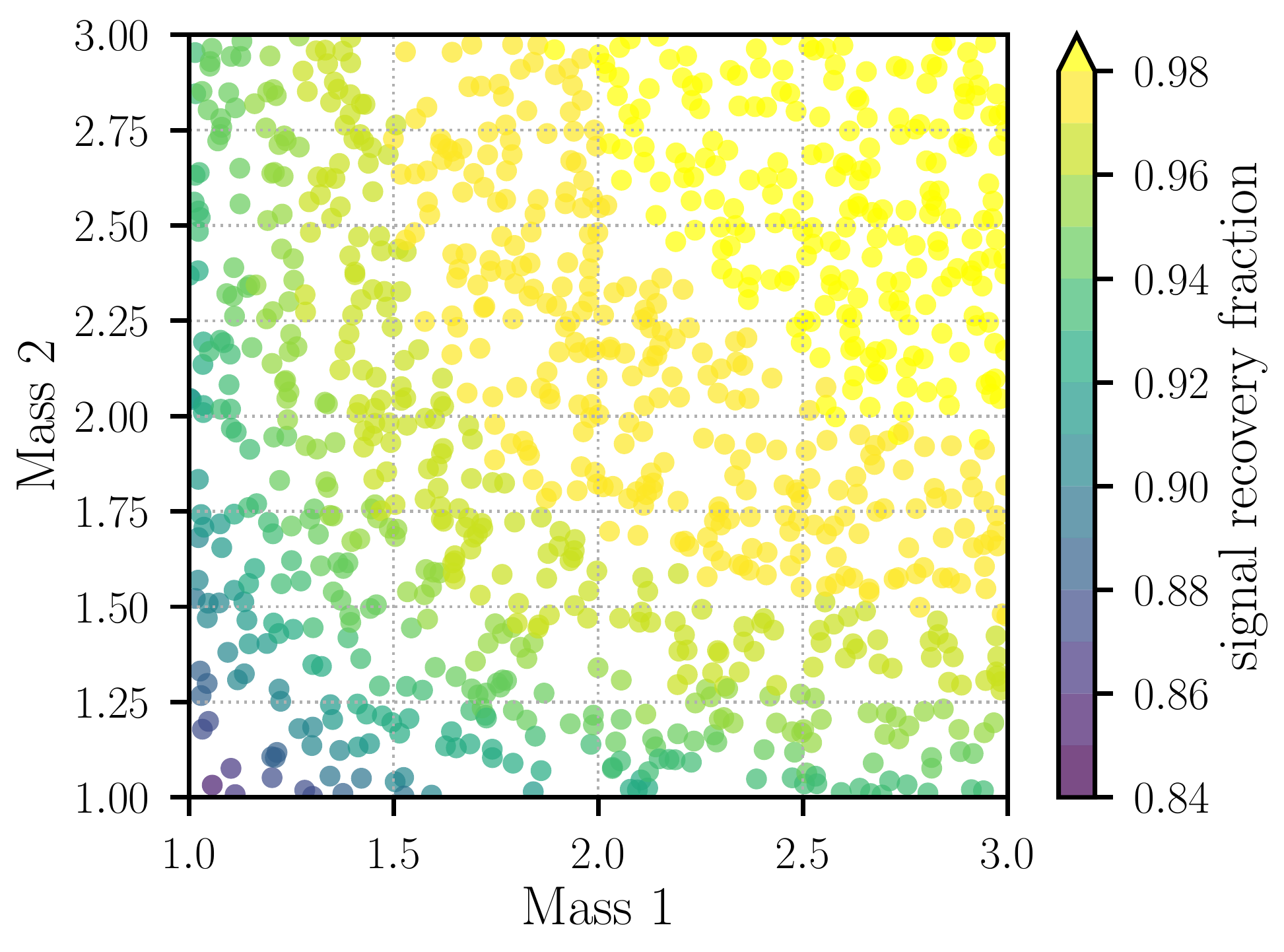}
    \hspace{0.02\linewidth}
    \includegraphics[width=0.49\linewidth]{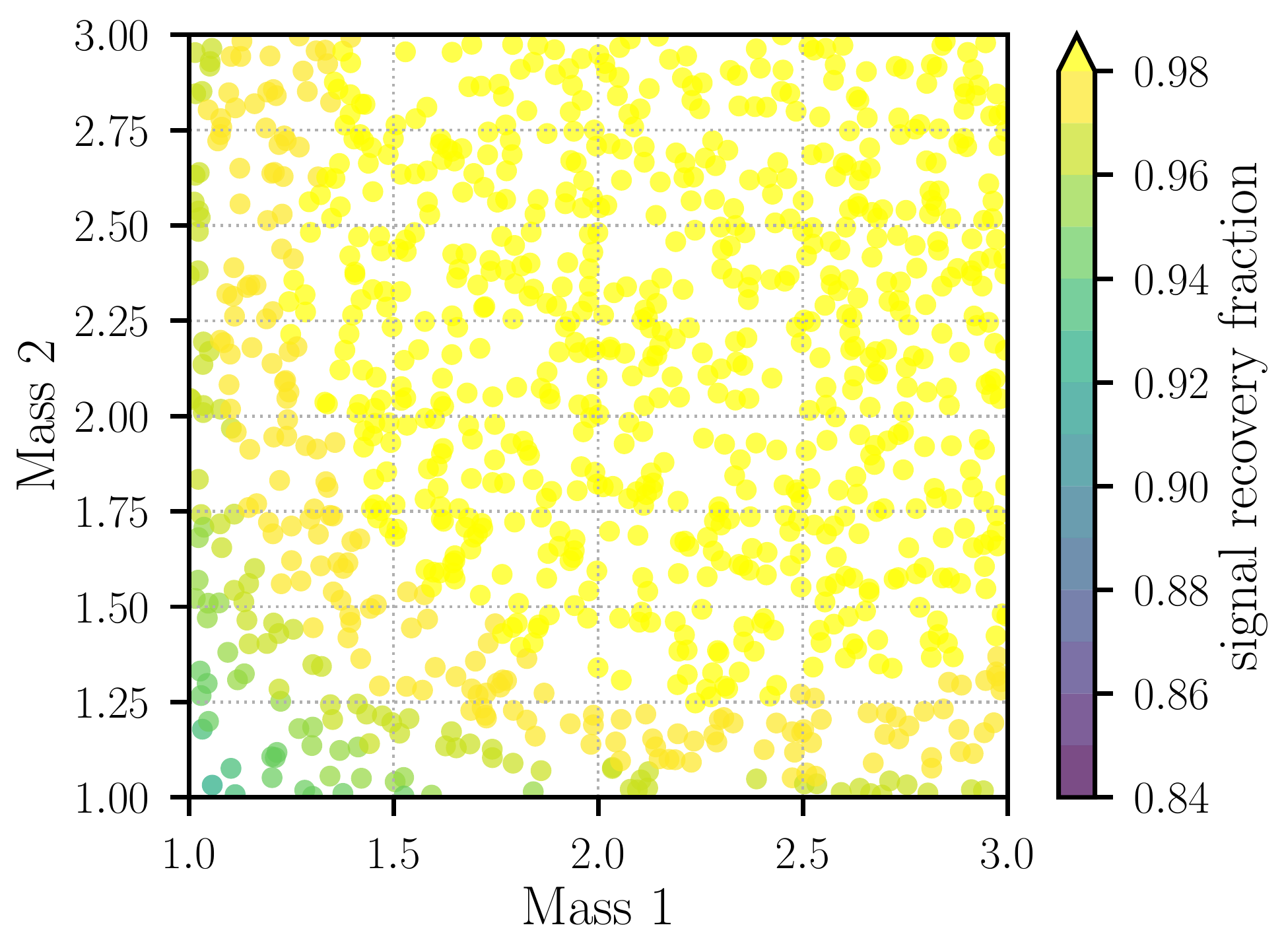}
  \end{minipage}
  \begin{minipage}[t]{1.0\linewidth}
    %\includegraphics[width=0.49\linewidth]{images/kerrtmpls_MS1Bsignals_noselfspin_srf.pdf}
    %\hspace{0.02\linewidth}
    %\includegraphics[width=0.49\linewidth]{images/kerrtmpls_MS1Bsignals_notidal_srf.pdf}
    \includegraphics[width=0.49\linewidth]{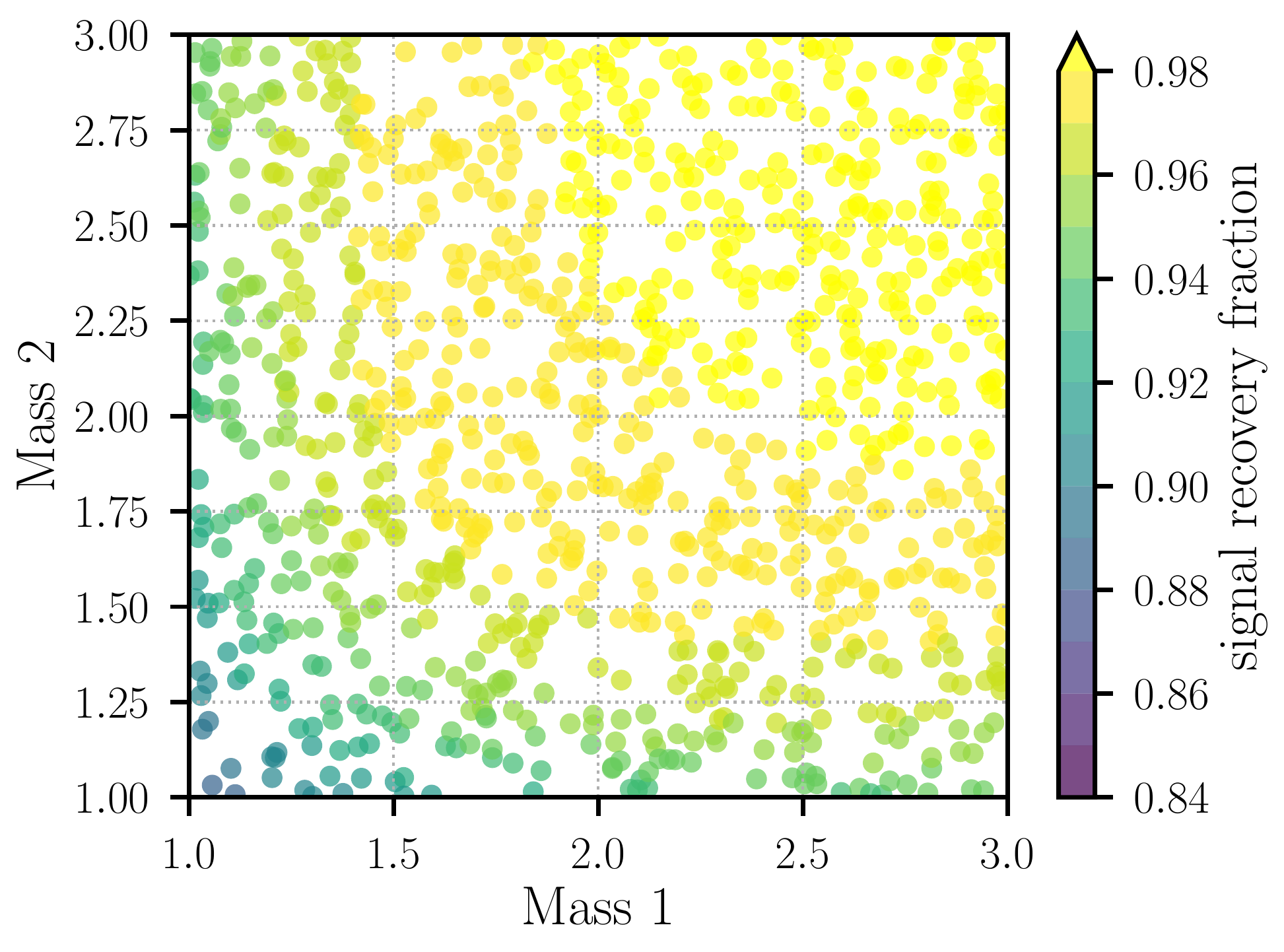}
    \hspace{0.02\linewidth}
    \includegraphics[width=0.49\linewidth]{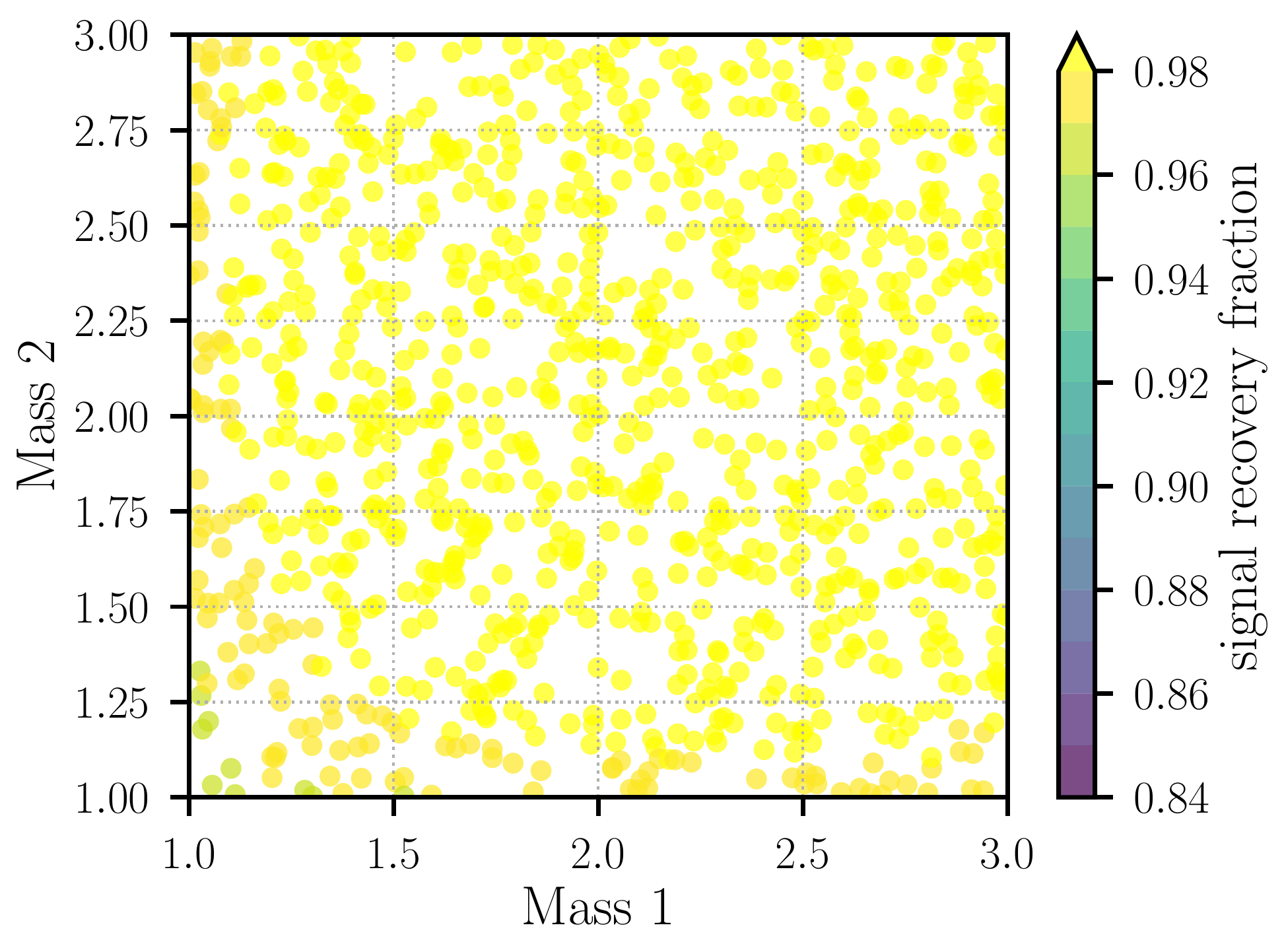}
  \end{minipage}
  \begin{minipage}[t]{1.0\linewidth}
    %\includegraphics[width=0.49\linewidth]{images/kerrtmpls_MS1Bsignals_match14.pdf}
    %\hspace{0.02\linewidth}
    %\includegraphics[width=0.49\linewidth]{images/kerrtmpls_SLYsignals_match14.pdf}
    \includegraphics[width=0.49\linewidth]{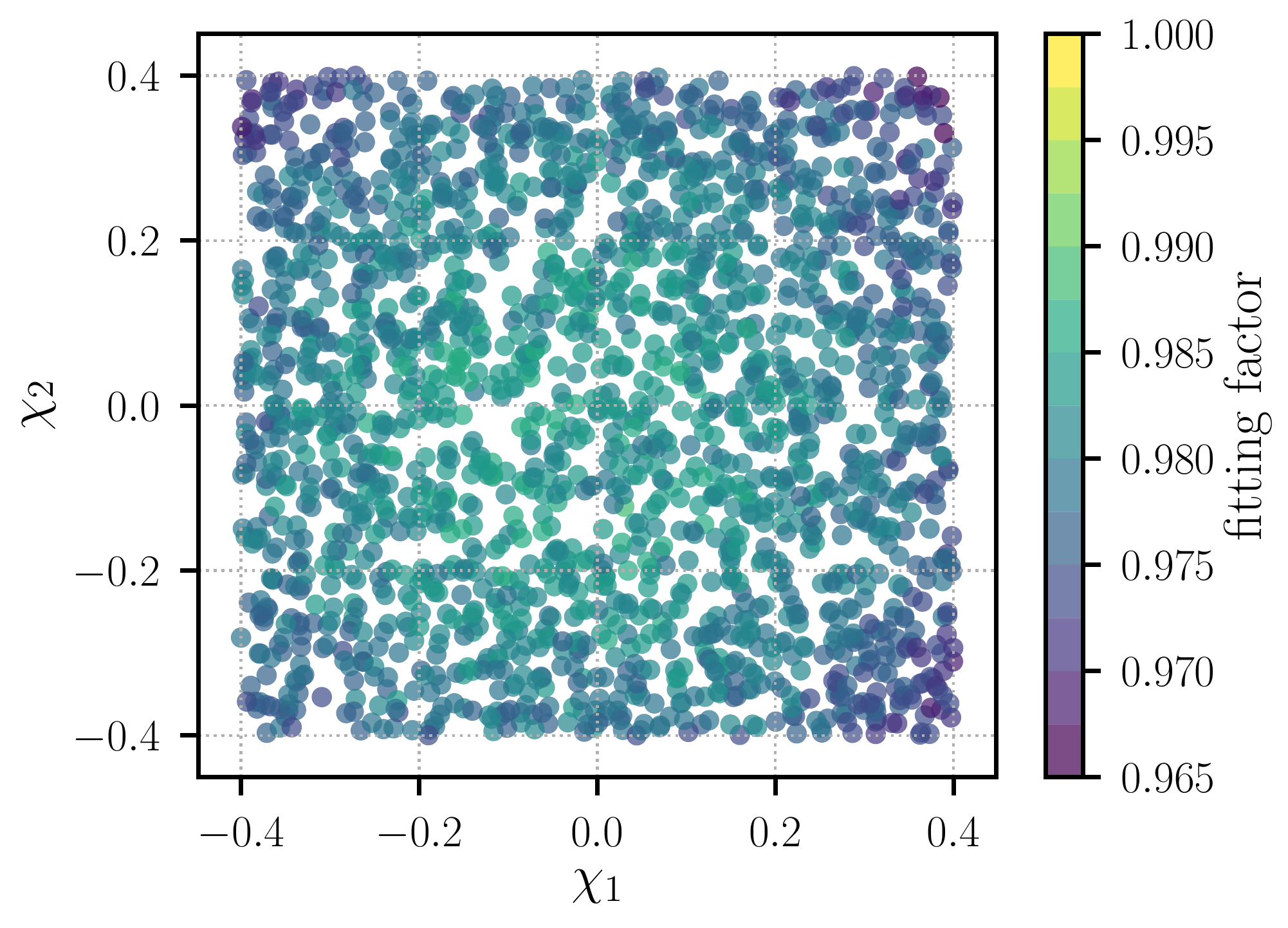}
    \hspace{0.02\linewidth}
    \includegraphics[width=0.49\linewidth]{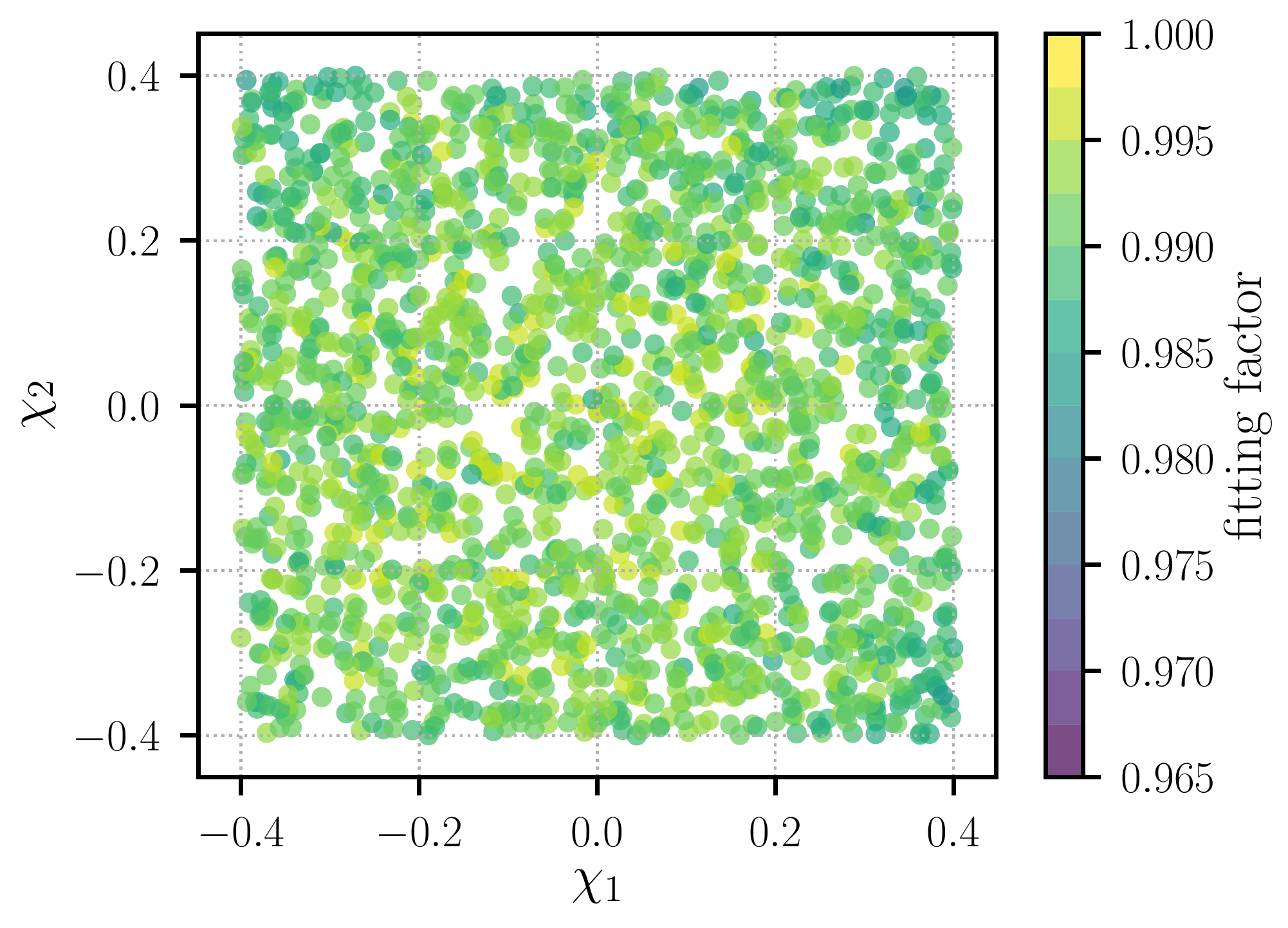}
  \end{minipage}
\caption{\label{fig:effectualness3}
      Top left (top right): Signal recovery fraction as a function of component masses for a population of
signals modelled using the MS1b (SLy) equations of state defined in section \ref{sec:waveforms} recovered
with a template bank containing signals modelled as if both bodies were Kerr black holes. Middle right:
Same as top left, using the MS1b equation-of-state except we set the self-spin term to a value of 1,
consistent with Kerr black holes. Middle left: Same as top left, using the MS1B equation-of-state except
we do not include the tidal deformation terms at 5 and 6PN in the signal model. Bottom left (bottom right):
Fitting factor as a function of component spins for signals modelled using the MS1b (SLy) equation-of-state
where both component masses are equal to 1.35 solar masses. All computations here assume the Advanced LIGO
zero-detuned, high-power sensitivity curve.
      }
\end{figure}

We then measure the signal-recovery fraction using a set of signals modeled using the MS1b and SLy equations of state
described earlier. These include terms for the spin-quadrupole, tidal terms and conditions on the termination frequency.
We again use the same distribution of component spins, source orientation and sky location as in figure~\ref{fig:effectualness1}.
The results of this simulation are shown in figure~\ref{fig:effectualness3}. We can see here that, especially at low masses,
ignoring equation-of-state effects results in a signal-recovery fraction as low as 86\% if assuming the MS1b equation of state,
and as low as 93\% if assuming the SLy equation of state. The biggest reduction is always 
at the lowest values of component masses where equation-of-state effects become most important.

To try to identify whether the loss in signal-recovery fraction shown in figure~\ref{fig:effectualness3} comes primarily
from the tidal deformation terms or from the spin-quadrupole terms
we  perform two additional runs with the MS1b equation of state. In one case we do not include the tidal terms in
the waveform, and in the second case we do not include the self-spin terms. These runs are shown in the middle
panels of figure~\ref{fig:effectualness3}. We can clearly see from these plots that the drop in signal-recovery
fraction when neglecting equation-of-state terms comes primarily from the tidal terms. This is expected given the degeneracy
between the spin-quadrupole terms and the component spins and mass ratio observed in figure~\ref{fig:effectualness2}.

From these results we therefore conclude that although the spin-quadrupole term can have a much larger effect
on an emitted gravitational waveform than tidal terms, variations in the spin-quadrupole term are strongly
degenerate with changes in the masses and component spins. Therefore the presence of spin-quadrupole terms
for highly spinning binary neutron star merger waveforms is unlikely to cause a reduction in the number of binary neutron star
signals that can be observed with Advanced LIGO. As already shown elsewhere~\cite{Cullen:2017oaz}, the presence of tidal
terms can cause a small reduction in the number of observed signals. These conclusions are consistent
with what is expected: tidal terms enter at much higher frequency and therefore cannot
be easily mimicked by variations in the masses or spins, as illustrated in figure~\ref{fig:info}. 

\section{Parameter recovery with equation-of-state dependent effects}
\label{sec:param_bias}

We have demonstrated in section~\ref{sec:faithfulness} that the spin-quadrupole terms, often ignored
in gravitational-wave data analysis, can have a significant effect on the emitted gravitational-wave signal
for systems containing rapidly spinning neutron stars. In section~\ref{sec:effectualness} we demonstrated
that while this effect is large, it is degenerate with changes in the mass ratio and component spins and
one would be able to observe such systems well using only waveforms modeling both bodies as Kerr black
holes. We find that the effect of tidal deformation, which only becomes relevant at higher frequencies,
is a larger problem when thinking of observing such systems than the spin-quadrupole term.

However, when observing a binary neutron-star system one will also want to measure the parameters of the system,
not only the neutron star equation-of-state but also the component masses and spins. Doing so will
allow a much better understanding of how neutron stars form, how binary systems evolve, and provide a census of
the properties of the astrophysical binary neutron star population.
In this section we try to answer two questions.
First, if the equation-of-state terms are not included in the waveform model being used to estimate the system's parameters,
or if an incorrect equation-of-state is assumed, by how much will the values of the parameters that are measured be biased?
Second, is it possible to measure the equation-of-state terms or
to test if a specific observation is more compatible with one equation-of-state compared to another.

\subsection{Methodology}
\label{ssec:param_bias_method}

To answer these questions we wish to evaluate 
\begin{equation}
 P(\xi_i | s, I) = \frac{P(\xi_i | I)}{P(s , I)} P(s | \xi_i, I),
\end{equation}
which defines the probability of a signal being present in the data with parameters given by $\xi_i$. When evaluated over
all $\xi_i$ this defines the probability-density function over the whole parameter space being considered. An introduction
of these concepts were given in section~\ref{sec:searchintro}.

In this work we use a novel technique for evaluating $P(\xi_i | s, I)$. For non-precessing
binary neutron star waveforms, there is only a weak coupling between the ``extrinsic'' parameters of the system---the sky-location,
distance, orientation and polarization phase---and the ``intrinsic'' parameters---the component masses, spins and the underlying
equation of state. Therefore we can make the approximation that analytically maximizing over the unknown extrinsic parameters of the
system is equivalent to marginalizing over these parameters. The validity of this approximation
is demonstrated in~\cite{Purrer:2015nkh}. It is then possible to randomly pick a very large
number of points, for each of the simulations described for this work we use $2.5 \times 10^{12}$ points,
from the underlying distribution given by $P(\xi_i, I)$---restricted to only intrinsic parameters---and
calculate $P(\xi_i | s, I)$ for all points assuming the ``zero-noise'' realisation as motivated in section~\ref{sec:searchintro}.
The Fisher Information Matrix is used to predict the match between each of the points being considered and the parameters
corresponding to the true source, using the implementation
discussed in \cite{Brown:2012qf}. For all points where the Fisher Information Matrix predicts that
the match is not negligibly small, $P(\xi_i | s, I)$ is calculated numerically using the \texttt{PyCBC}
software package~\cite{Canton:2014ena,Usman:2015kfa,alex_nitz_2017_1058970}. At all other points $P(\xi_i | s, I)$ is assumed to be 0.
In this way we can rapidly evaluate $P(\xi_i | s, I)$ for the ``zero-noise'' realization.

\subsection{Results}
\label{ssec:param_bias_selfspinonly}

\begin{figure}
  \centering
%  \begin{minipage}[t]{1.0\linewidth}
%    \includegraphics[width=0.49\linewidth]{images/saff_1p351p35_spin_0p350p35_ssonly.pdf}
%    \hspace{0.02\linewidth}
%    \includegraphics[width=0.49\linewidth]{images/saff_1p351p35_spin_0p350p35_snr12_ssonly.pdf}
%  \end{minipage}
%  \begin{minipage}[t]{1.0\linewidth}
%    \includegraphics[width=0.49\linewidth]{images/saff_1p351p35_spin_0p00p0_ssonly.pdf}
%    \hspace{0.02\linewidth}
%    \includegraphics[width=0.49\linewidth]{images/saff_1p41p1_spin_0p350p35_ssonly.pdf}
%  \end{minipage}
  \begin{minipage}[t]{1.0\linewidth}
    \includegraphics[width=0.49\linewidth]{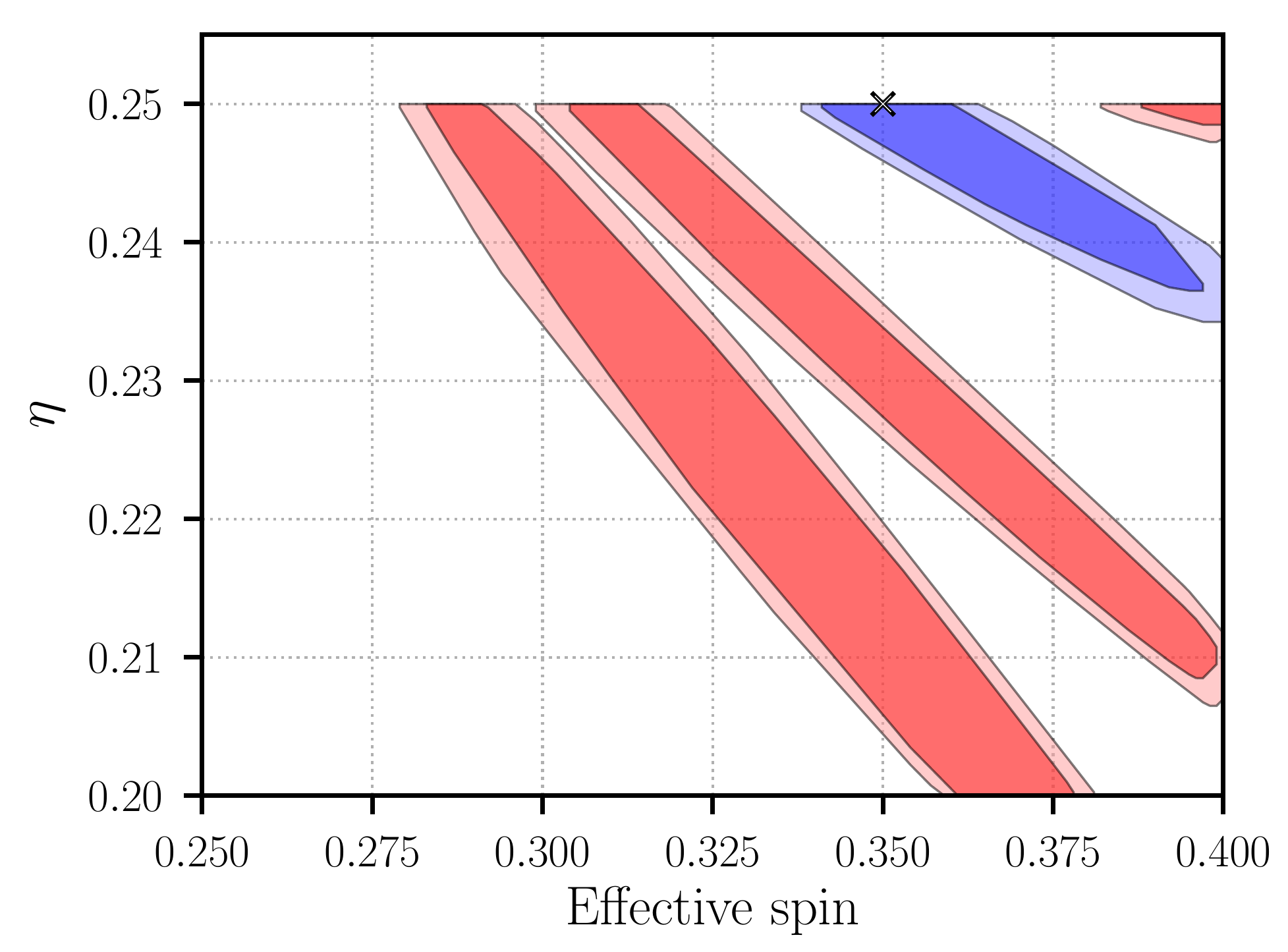}
    \hspace{0.02\linewidth}
    \includegraphics[width=0.49\linewidth]{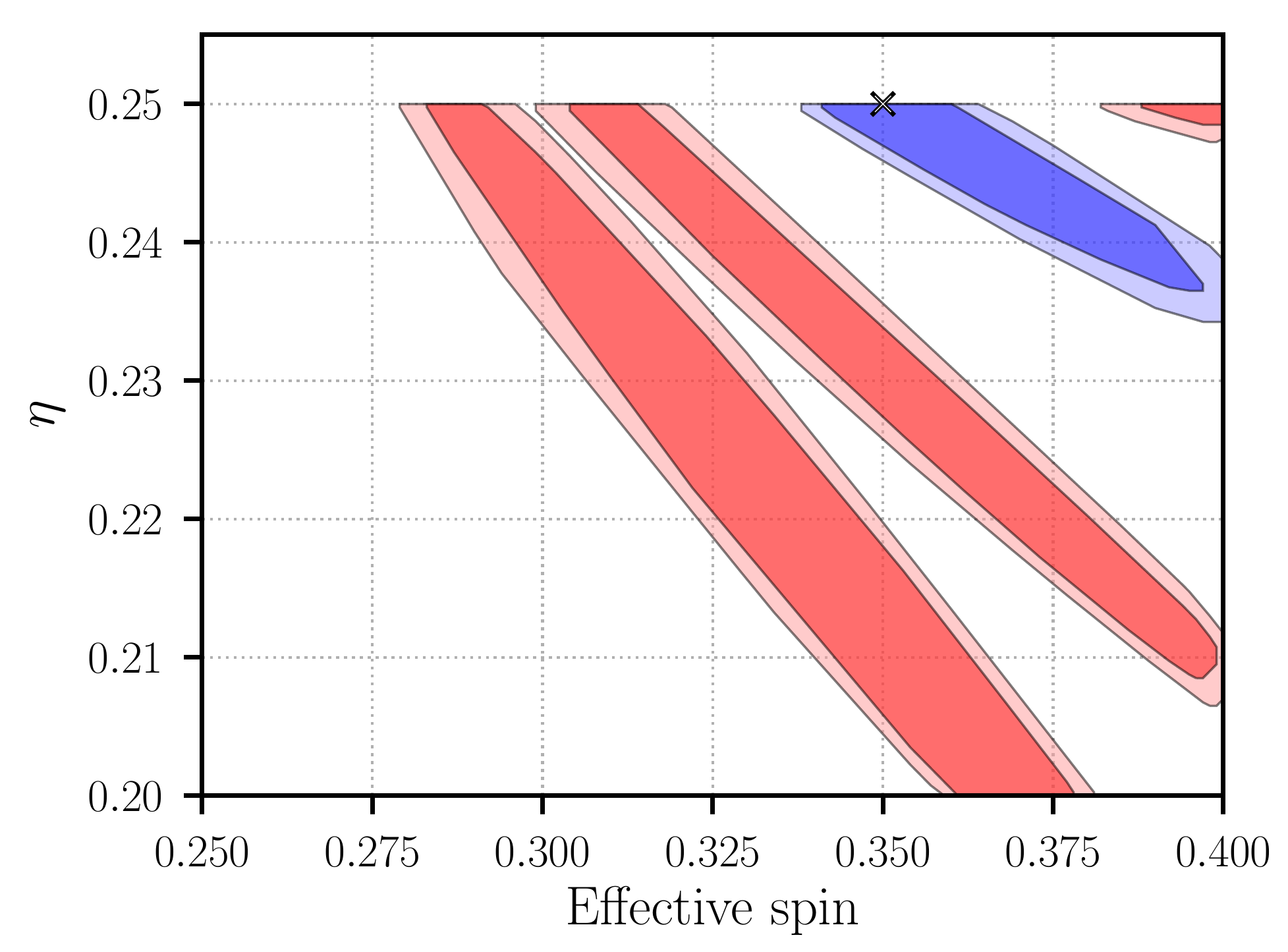}
  \end{minipage}
  \begin{minipage}[t]{1.0\linewidth}
    \includegraphics[width=0.49\linewidth]{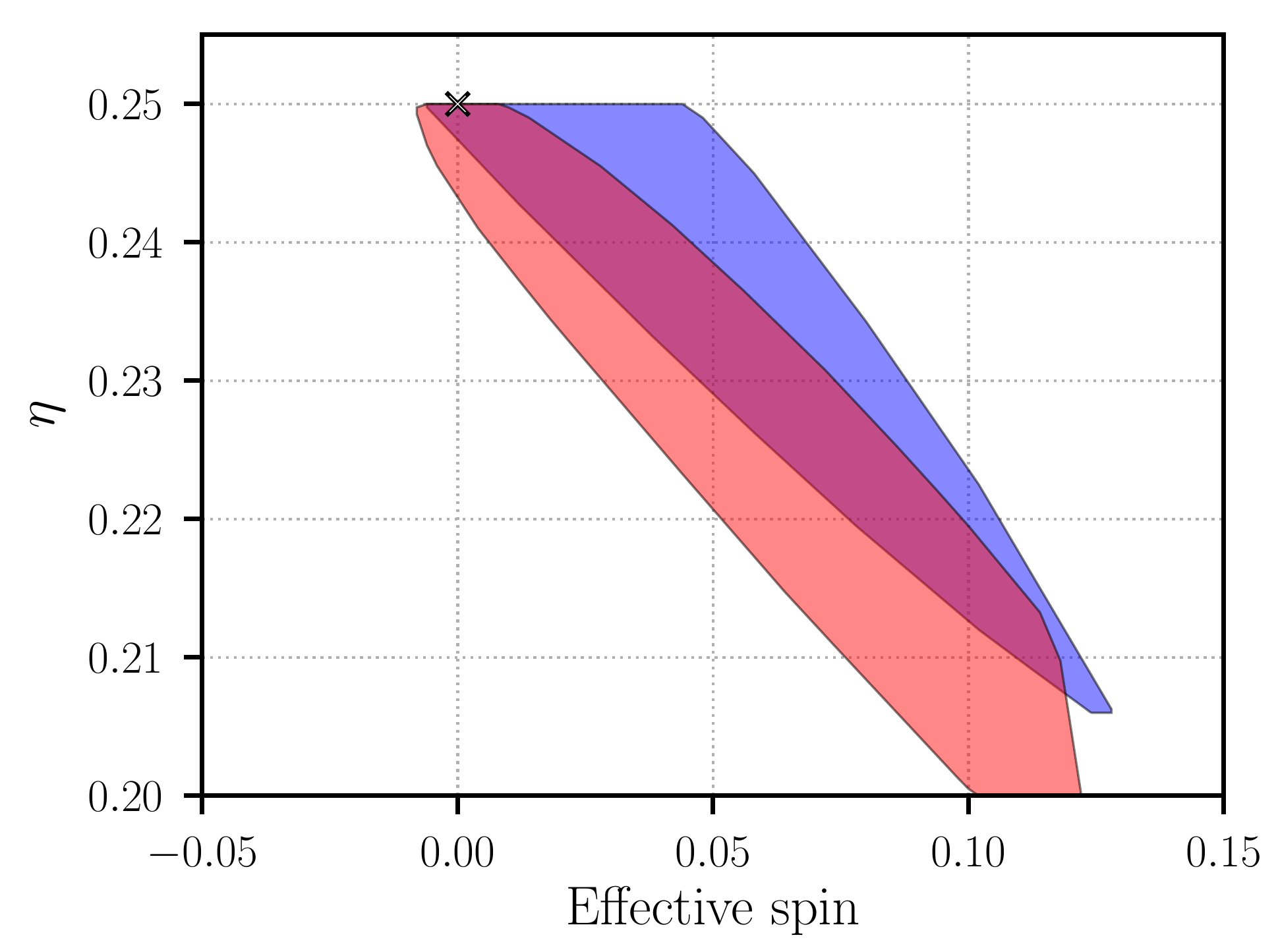}
    \hspace{0.02\linewidth}
    \includegraphics[width=0.49\linewidth]{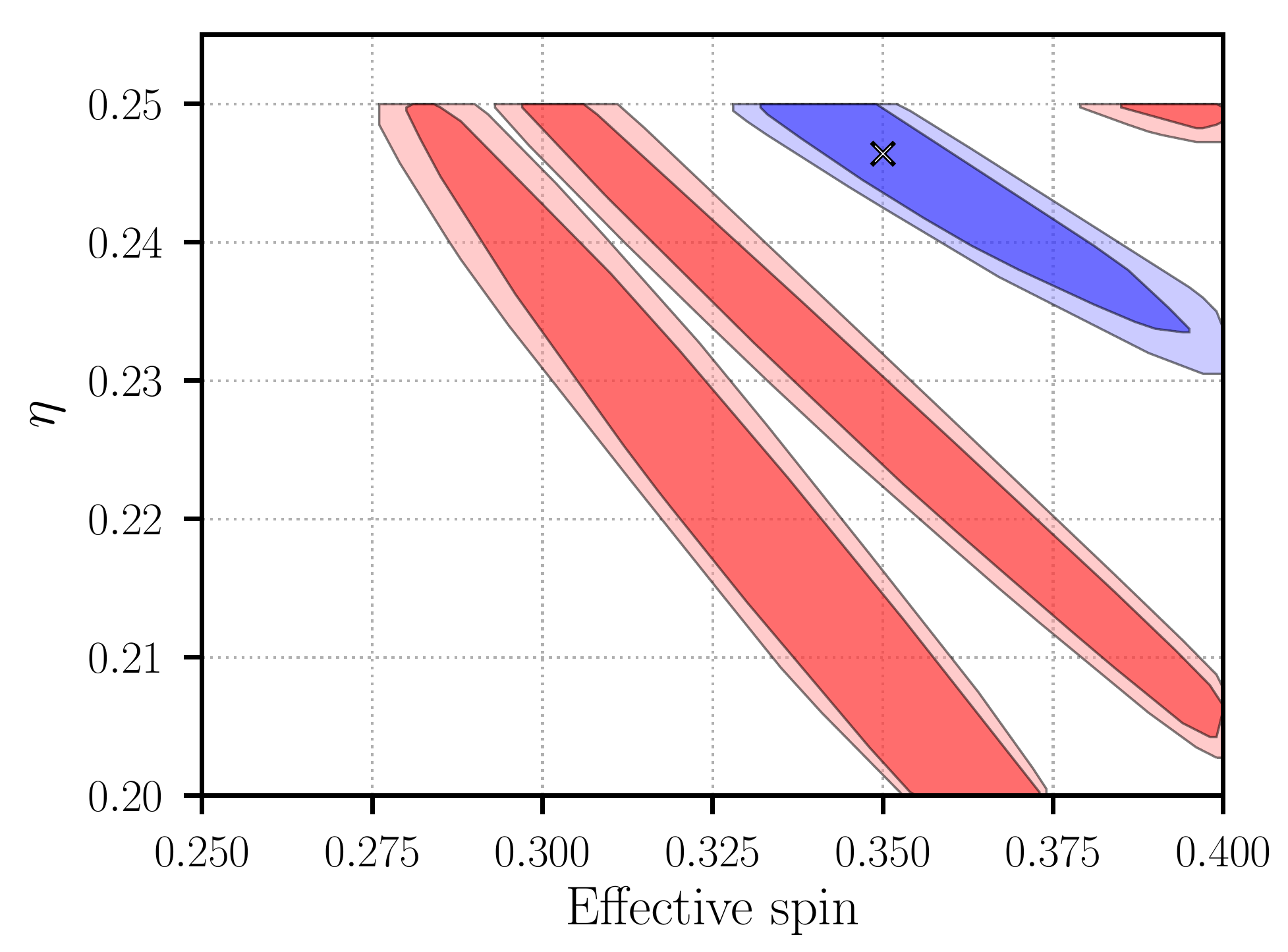}
  \end{minipage}
      \caption{\label{fig:saff1}
      Two-dimensional marginalized posterior probability distribution of effective spin and the symmetric
      mass ratio, $\eta$, for the four simulations listed in table~\ref{tab:saff1}. The top left panel
      corresponds to ID 1, the top right to ID 2, the bottom left to ID 3 and the bottom right to ID 4.
      The blue shaded region denotes the 99\% and 99.99\% confidence region when using the correct value
      of the spin-quadrupole term, which is assumed to be $\mathcal{Q}=8$ here for all simulations. The
      red shaded regions denote the 99\% and 99.99\% when assuming incorrect values of the spin-quadrupole
      term, shown here are results for $\mathcal{Q}=1,4$ and $12$. As simulation ID 3 is modelled with both
      component spins of 0, the bias here is much smaller than for other cases. Here, for clarity, we only show the 99\%
      confidence regions for $\mathcal{Q}=8$ and $\mathcal{Q}=1$.
      }
\end{figure}

\begin{table}[tbp]
  \centering
  \begin{tabular}{c|c|c|c|c|c|c|c|c|c}
    ID & $m_1$ & $m_1$ & $\chi_1$ & $\chi_2$ & $\rho$ & $\mathcal{L}_{\mathcal{Q}=1}$
    &  $\mathcal{L}_{\mathcal{Q}=4}$ &  $\mathcal{L}_{\mathcal{Q}=8}$ &  $\mathcal{L}_{\mathcal{Q}=12}$ \\ \hline
    1 & 1.35 & 1.35 & 0.35 & 0.35 & 25 & 0.47 & 1.97 & 1.0 & $1\times10^{-14}$\\
    2 & 1.35 & 1.35 & 0.35 & 0.35 & 12 & 3.86 & 3.44 & 1.0 & $3\times10^{-5}$ \\
    3 & 1.35 & 1.35 & 0. & 0. & 25 & 0.76 & 1.56 & 1.0 & 2.20 \\
    4 & 1.4 & 1.1 & 0.35 & 0.35 & 25 & 0.33 & 1.18 & 1.0 & $1\times10^{-8}$ \\
  \end{tabular}
    \caption{\label{tab:saff1} Parameters, and marginalized likelihood, for the runs plotted in figure~\ref{fig:saff1}.
    $m_1$ and $m_2$ denote the two component masses in units of component masses. $\chi_1$ and $\chi_2$ denote the
    two component masses. $\rho$ denotes the signal-to-noise ratio of the signal. $\mathcal{L}_{\mathcal{Q}=x}$ gives
    the marginalized likelihood when searching for the signal assuming the spin-quadrupole term, $\mathcal{Q}$, is $x$
    for both bodies. In all cases the signal is modelled using $\mathcal{Q}=8$
    and the marginalized likelihoods are  normalized so that $\mathcal{L}_{\mathcal{Q}=8} = 1$.
    Then the ratios of these values give
    the relative posterior likelihood between the various models. All simulations here assume the Advanced LIGO zero-detuned,
    high-power noise curve.}
\end{table}

\begin{figure}
  \centering
%  \begin{minipage}[t]{1.0\linewidth}
%    \includegraphics[width=0.49\linewidth]{images/saff_1p351p35_spin_0p350p35_fulltidal.pdf}
%    \hspace{0.02\linewidth}
%    \includegraphics[width=0.49\linewidth]{images/saff_1p351p35_spin_0p350p35_snr12_fulltidal.pdf}
%  \end{minipage}
%  \begin{minipage}[t]{1.0\linewidth}
%    \includegraphics[width=0.49\linewidth]{images/saff_1p351p35_spin_0p00p0_fulltidal.pdf}
%    \hspace{0.02\linewidth}
%    \includegraphics[width=0.49\linewidth]{images/saff_1p41p1_spin_0p350p35_fulltidal.pdf}
%  \end{minipage}
  \begin{minipage}[t]{1.0\linewidth}
    \includegraphics[width=0.49\linewidth]{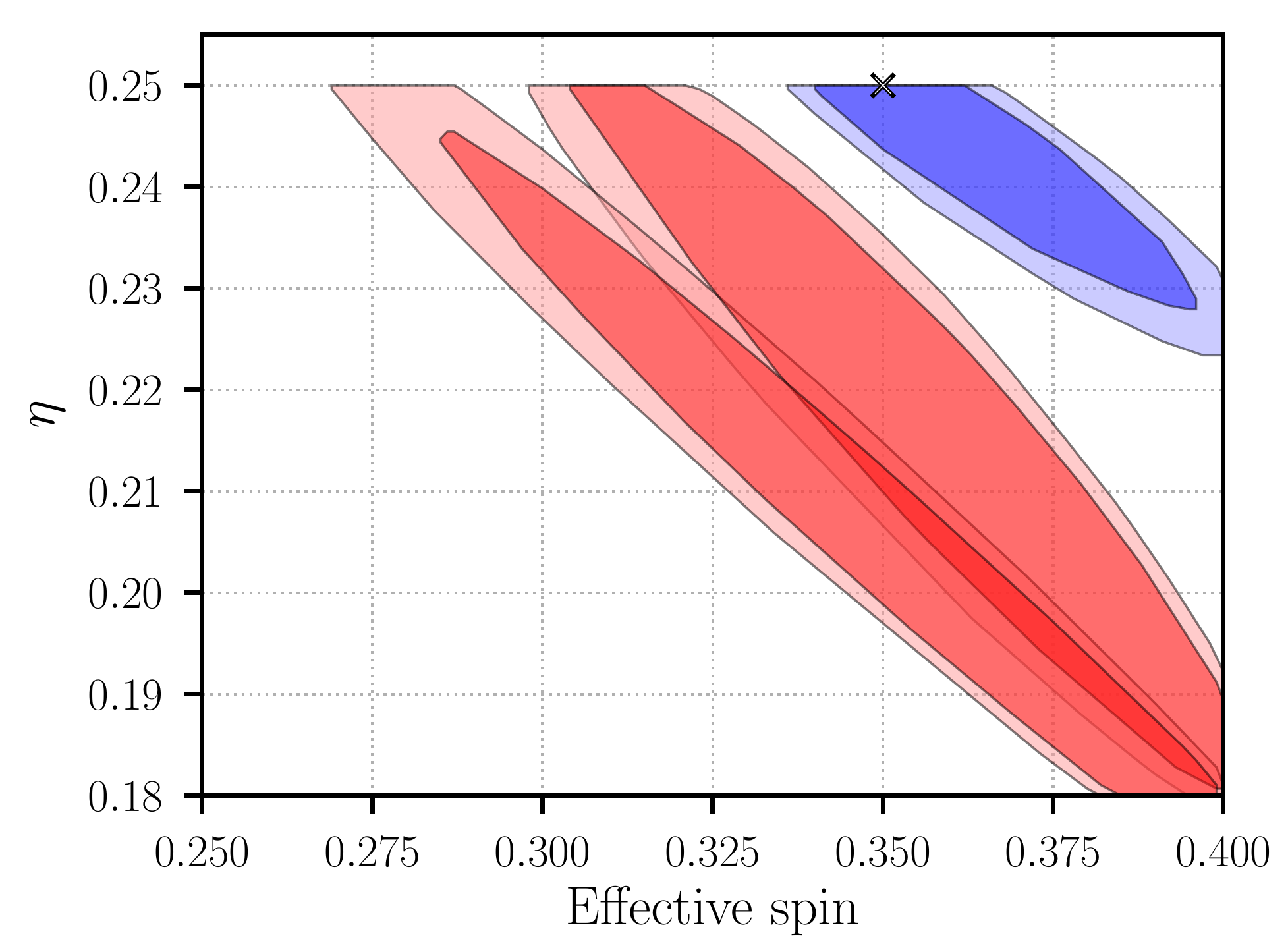}
    \hspace{0.02\linewidth}
    \includegraphics[width=0.49\linewidth]{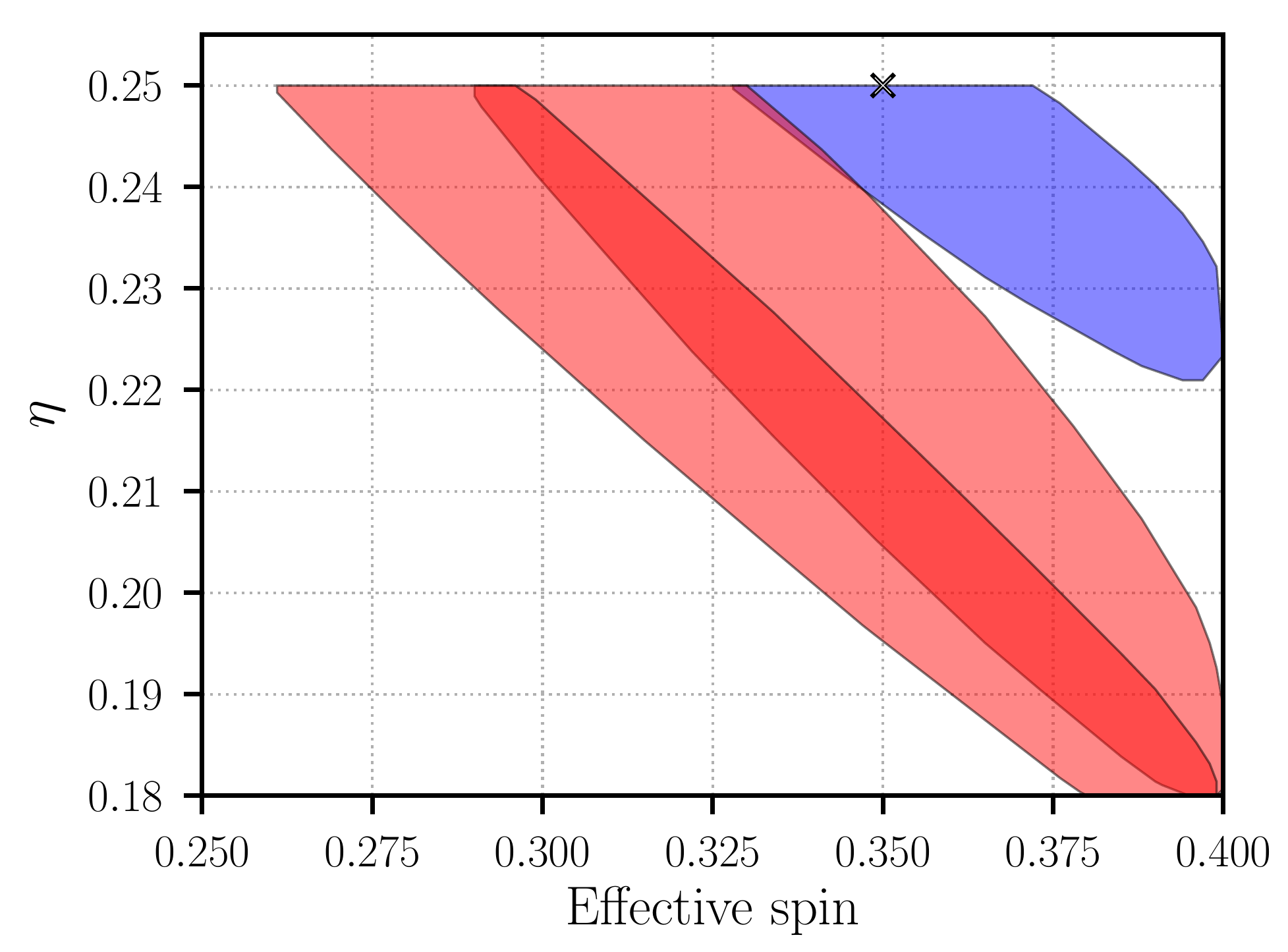}
  \end{minipage}
  \begin{minipage}[t]{1.0\linewidth}
    \includegraphics[width=0.49\linewidth]{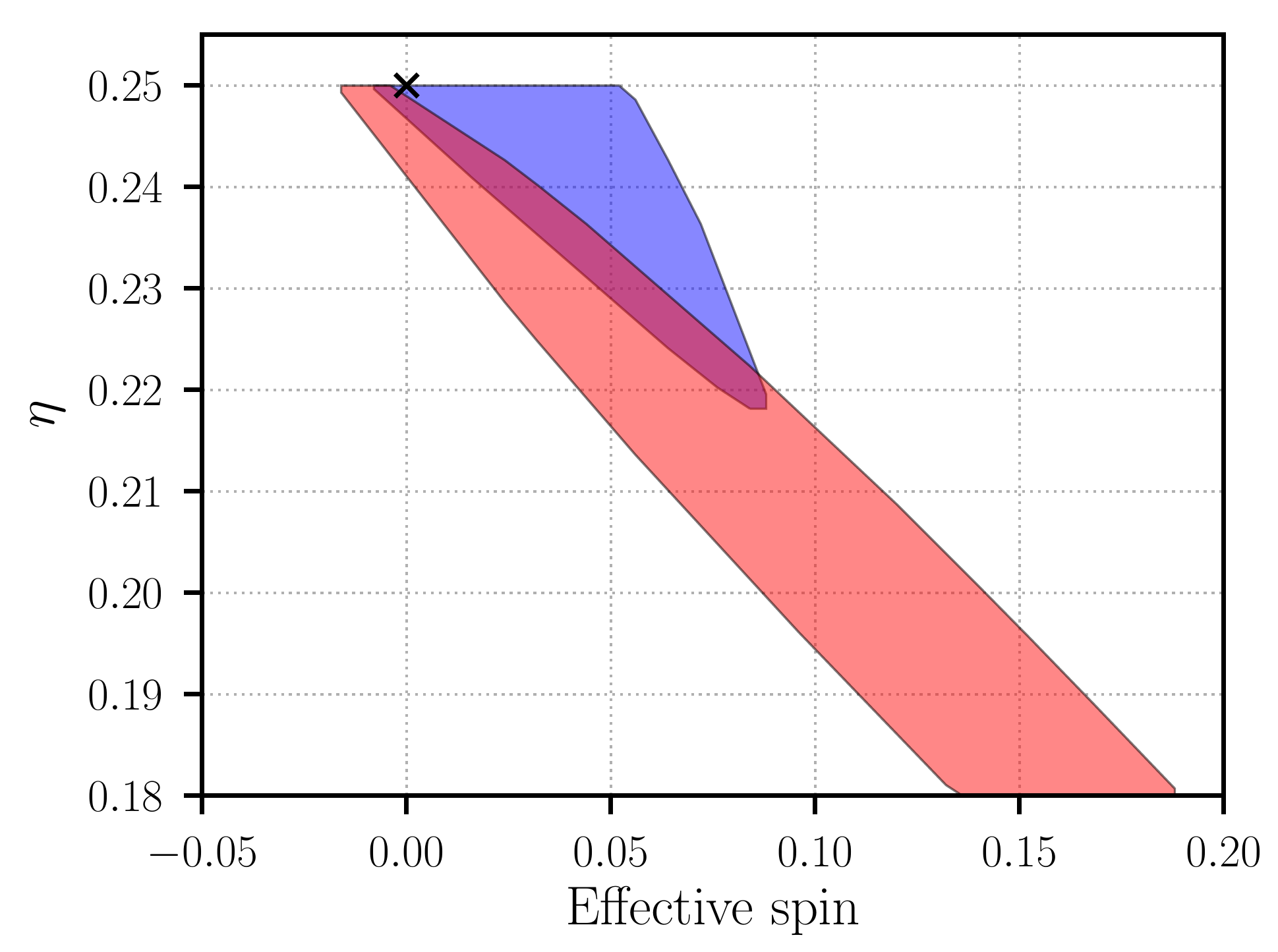}
    \hspace{0.02\linewidth}
    \includegraphics[width=0.49\linewidth]{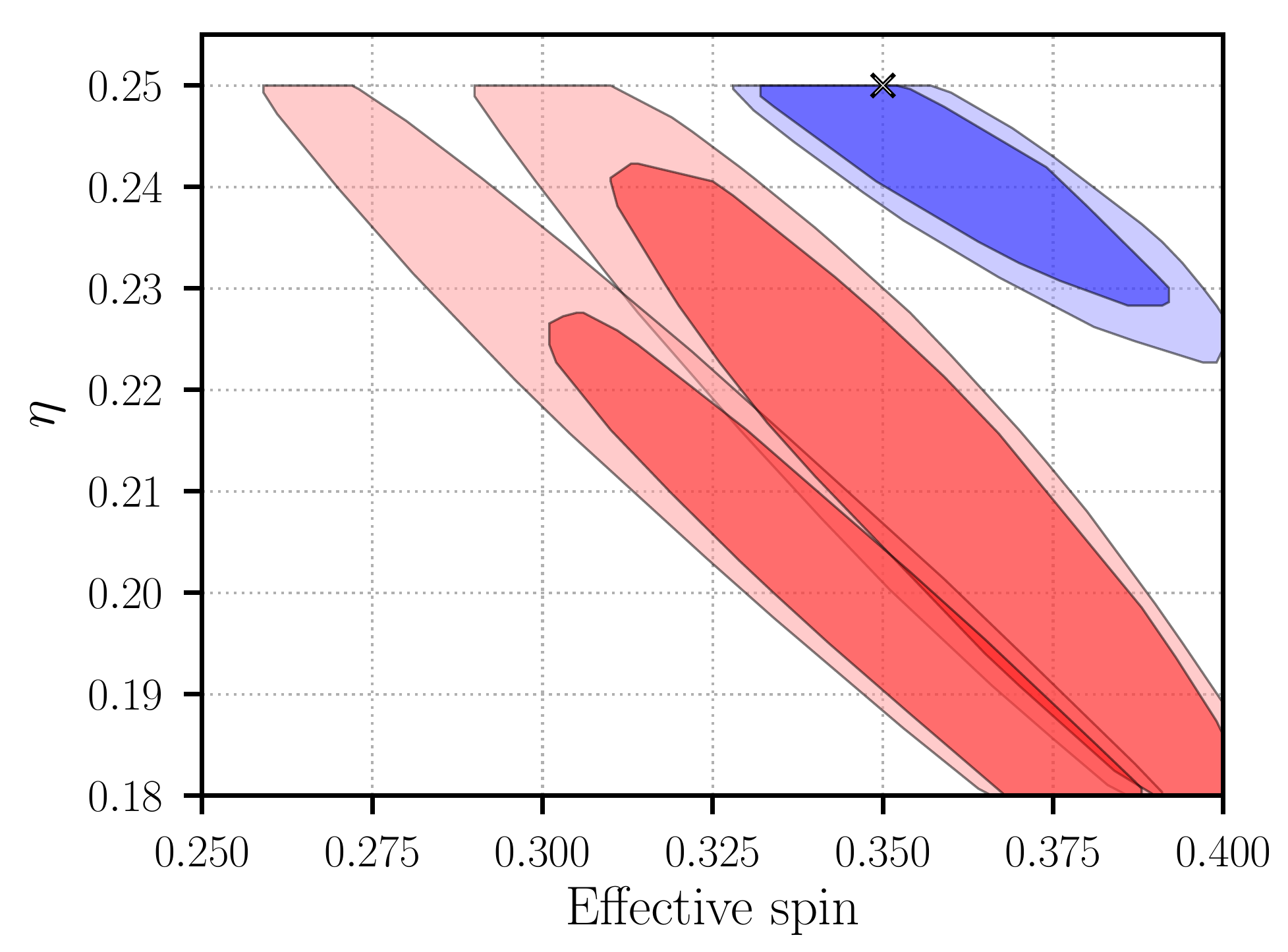}
  \end{minipage}
      \caption{\label{fig:saff2}
      Two-dimensional marginalized posterior probability distribution of effective spin and the symmetric
      mass ratio, $\eta$, for the four simulations listed in table~\ref{tab:saff2}. The top left panel
      corresponds to ID 5, the top right to ID 6, the bottom left to ID 7 and the bottom right to ID 8.
      The blue shaded region denotes the 99\% and 99.99\% confidence region when using the correct 
      equation-of-state, which is MS1b here for all simulations. The
      red shaded regions denote the 99\% and 99.99\% when assuming incorrect values of the equation-of-state
      term, shown here are results for SLy, and when assuming both bodies are low mass black holes.
      As simulation ID 7 is modelled with both
      component spins of 0, the bias here is much smaller than for other cases. Here, for clarity, we only show the 99\%
      confidence regions for MS1b, and two Kerr black holes. We also show only the 99\% confidence regions
      in simulation ID 6.
      }
\end{figure}

\begin{table}[tbp]
  \centering
  \begin{tabular}{c|c|c|c|c|c|c|c|c}
    ID & $m_1$ & $m_1$ & $\chi_1$ & $\chi_2$ & $\rho$ &
    $\mathcal{L}_{\mathrm{EOS}=\mathrm{MS1b}}$ &  $\mathcal{L}_{\mathrm{EOS}=\mathrm{SLy}}$ &
    $\mathcal{L}_{\mathrm{EOS}=\mathrm{Kerr}}$ \\ \hline
    5 & 1.35 & 1.35 & 0.35 & 0.35 & 25 & 1.0 & 0.0019 & $1.7\times10^{-5}$\\
    6 & 1.35 & 1.35 & 0.35 & 0.35 & 12 & 1.0 & 0.75 & 0.30 \\
    7 & 1.35 & 1.35 & 0. & 0. & 25 & 1.0 & 0.18 & 0.011 \\
    8 & 1.4 & 1.1 & 0.35 & 0.35 & 25 & 1.0 & 0.00025 & $5\times10^{-7}$ \\
  \end{tabular}
    \caption{\label{tab:saff2} Parameters, and marginalized likelihood, for the runs plotted in figure~\ref{fig:saff2}.
    $m_1$ and $m_2$ denote the two component masses in units of component masses. $\chi_1$ and $\chi_2$ denote the
    two component masses. $\rho$ denotes the signal-to-noise ratio of the signal. $\mathcal{L}_{\mathrm{EOS}=x}$ gives
    the marginalized likelihood when searching for the signal assuming an equation-of-state of $x$ for both bodies, either
    MS1b, SLy or that the body is a Kerr black hole. In all cases the signal is modelled using the MS1b equation of state
    and the marginalized likelihoods are
    normalized so that $\mathcal{L}_{\mathrm{EOS}=MS1b} = 1$. Then the ratios of these values give
    the relative posterior likelihood between the various models. All simulations here assume the Advanced LIGO zero-detuned,
    high-power noise curve.}
\end{table}

We begin by exploring the parameter bias that occurs if searching for binary neutron star systems using waveforms where the
value of the
spin-quadrupole term differs from the signal we are looking for. We do this for systems that have a number of different
values of masses and component spins. In all cases the signal we are looking for is assumed to have a spin-quadrupole
value of ${\cal Q}=8$ on both bodies, and we try to recover this signal assuming ${\cal Q}=1, 4, 8$ or $12$ on both
neutron stars. This allows us to understand how the bias that will be present in measuring parameters varies
as the error on the spin-quadrupole value changes.
For this simulation we neglect tidal terms and use a termination frequency
corresponding to the binary black hole ``ISCO''-criterion.
The results of this are shown in figure~\ref{fig:saff1}. Here we show results for four different systems, with the details of those
systems given in table~\ref{tab:saff1}. We use a dimensionless spin of 0.35 to model the signal in many
cases in figure~\ref{fig:saff1}. While the bias is largest
for systems when the binary neutron star spins are large, there is a visible bias even when the source has no spin on either body. The reason for this
is that the signal from a non-spinning binary-neutron star system can match well with a system with non-zero spins. As the signal
from spinning systems is altered by the value of the spin-quadrupole term we still observe a bias if we allow high-spinning
systems in our prior. We also show the posteriors marginalized over the full parameter space for all cases in table~\ref{tab:saff1}.
Here the majority of information is coming from the priors, and boundaries that we have placed on the parameter space, rather
than from the data. For example we notice that in all cases with component spins of 0.35 the $\mathcal{Q}=12$ case is strongly
disfavored. This is because we have assumed that the prior probability on component spins is flat up to a value of 0.4
and then drops immediately to 0. For $\mathcal{Q}=12$ we require spins larger than 0.4 to match well to the assumed signal model,
which are not permitted, and therefore strongly disfavored. Likewise, in many cases values of $\mathcal{Q}=4$ are favored above
the correct value $\mathcal{Q}=8$, again this is because the 2-dimensional probability plots shown in figure~\ref{fig:saff1}
intersect the boundary of the parameter space to a greater degree with $\mathcal{Q}=8$ than with $\mathcal{Q}=4$. In short
we are not able to measure the value of $\mathcal{Q}$ in any of these simulations.

An additional test of the parameter bias is shown in figure~\ref{fig:saff2} and table~\ref{tab:saff2}.
In contrast to figure~\ref{fig:saff1}, here we
use the fits to the various equation of states described in section~\ref{sec:waveforms} to evaluate the parameter
bias that would be present if we search for a binary-neutron system described by one equation-of-state using waveforms
modelled by another. In addition to the spin-quadrupole term, these waveforms include equation-of-state and mass-dependent
terms describing the tidal deformation and the termination frequency. In these results we model the astrophysical signal
using the MS1b equation-of-state and show results searching for this signal using the MS1b and SLy equations-of-state as
well as results assuming that both compact bodies are Kerr black holes. Other than that we use the same priors as described
above for figure~\ref{fig:saff2}. As with figure~\ref{fig:saff2} we see that assuming the incorrect equation-of-state when
measuring source parameters can lead to a significant bias. In all cases we can see that the permitted range of mass ratios
is much more constrained when using the correct MS1b equation-of-state than when using SLy or assuming two Kerr black holes.
This is because the tidal terms become increasingly large on the smaller body as the mass ratio increases for MS1b and this
breaks the mass-ratio and spin degeneracy much more easily. We also show the marginalized likelihoods for the various cases
in table~\ref{tab:saff2}. We can see that for the two high-spin, high signal-to-noise-ratio cases the correct equation of state
is strongly favoured over SLy or the Kerr black hole model. For the non-spinning, high signal-to-noise ratio case the discrimination
power is much weaker than for the spinning cases. This implies that measuring the equation-of-state for a spinning binary neutron
star will be much easier than for a non-spinning binary neutron star. Finally, when reducing the signal-to-noise ratio to 12, as
might be expected for many of the binary-neutron star merger detections, we find that it is not possible to distinguish between the various
equations of states.

\section{Redshift effects}
\label{sec:redshift}

It has been pointed out in previous works that measuring the tidal deformability terms $\Lambda_i$ precisely offers a way
to directly measure the object's masses and hence the redshift of the source~\cite{Messenger:2011gi}. The reason for this is that gravitational-wave signals are redshifted in an analogous
way to electromagnetic signals. In the case of binary-black hole mergers the redshift effect is entirely degenerate with
the total mass of the system. That is to say that the gravitational-wave signal observed from a redshifted source will
appear exactly the same as that from a non-redshifted source with larger masses by a factor of $(1+z)$, where $z$ is the redshift. However, in the case of binary neutron
stars the tidal deformability breaks this degeneracy. The reason is that $\Lambda_i$ depends on the source-frame
mass as $\Lambda_i \sim m_i^{-5}$, where a more detailed approximation involves a linear expansion around a reference
value $\Lambda(m=1.4M_\odot)$. Therefore one can simultaneously measure the redshifted mass and the tidal deformability
of a system, and if the relationship between neutron-star mass and tidal deformability is well
understood, one can then determine the redshift of the system because the tidal terms will involve an explicit factor of $(1+z)^5$.

Similarly, the presence of the quadrupole-monopole term also breaks the degeneracy between total mass and redshift, and a precise
measurement of the quadrupole-monopole term could likewise allow a determination of a source's redshift.
The quadrupole parameter scales with the source-frame mass as ${\cal Q}_i\sim m_i^{-1}$, and an analogous reasoning as for
the tidal terms applies. However, the results
in this paper demonstrate that the quadrupole-monopole term is not measurable to the same accuracy as the tidal deformation
term, as changes in the quadrupole-monopole term are degenerate with changes in the component spins and mass ratio. Therefore
the ability to measure the redshift of a binary-neutron star system will still depend on the ability to measure the tidal
deformation of inspiraling neutron stars. 

\section{Conclusion}
\label{sec:conclusion}

In this work we have explored the affects of the neutron-star equation-of-state in binary neutron star observations
with a particular focus on spinning binary neutron stars and the spin-quadrupole term that is often ignored, as it was
in~\cite{TheLIGOScientific:2017qsa}.
We have explored the overall distinguishability of waveforms as a function of equation-of-state related terms and found
that the spin-quadrupole term has a much larger effect than the tidal-deformability for highly spinning neutron star systems,
although both terms are potentially measurable if all other parameters about the system are known. We have explored whether
the equation-of-state would have any affect on our ability to observe binary neutron-star mergers, where it is commonly assumed
that the compact objects are both black holes. We found that the tidal deformability can lead to a small reduction in the number
of observed binary neutron-star systems, as reported in~\cite{Cullen:2017oaz}, but found that the spin-quadrupole term is
largely degenerate with the component spins and mass ratio. Therefore, ignoring this will not result in a reduction in the
detection rate other than at the boundaries of the searched parameter space. We have explored the bias in recovered source parameters
that can be expected if making incorrect assumptions about the neutron-star equation-of-state and have found that the recovered values
of the component spins and mass ratio can be significantly biased. We have also explored the measurability of various equation-of-state
terms, finding that the spin-quadrupole term alone cannot be easily measured with Advanced LIGO,
but that combined with the effects of the tidal deformability it will be possible to rule out specific equations-of-state, especially
in the case that the neutron stars' component spins are large.

As always with investigations of this type, our results are only as good as the waveform model used. The Post-Newtonian waveform
model we used has two caveats, first that there may be uncontrolled systematic errors in the phasing model because the Post-Newtonian approximation breaks down in the relativistic regime close to the merger, and second because the waveform stops abruptly at a given frequency, whereas a real binary neutron star source might be expected to have a complex post-merger
structure. However, for the case of exactly equal mass systems the TaylorF2 inspiral waveforms give similar results as more sophisticated waveform models~\cite{Sennett:2017etc}, and the post-merger signals are expected to be at much larger frequencies than those that Advanced LIGO can observe. 
Thus, we expect our broad conclusions to remain unchanged in more realistic contexts. While work to improve waveform models in the binary-neutron star regime will
be important to remove the potential for systematic errors in gravitational-wave observations, we believe that the results and conclusions that we derive in this manuscript will remain largely unaltered. 

With many additional binary neutron-star mergers being expected to be observed by Advanced LIGO and Advanced Virgo in the coming
years it will be interesting to apply these methods also on real data and try to measure directly the equation-of-state of neutron
stars, as already explored in~\cite{TheLIGOScientific:2017qsa}. However, as such works are allowing for the possibility
of very high spin neutron-star systems, neglecting the spin-quadrupole terms \emph{will} lead to biased results, so we emphasize
the importance of including this term both in waveform modelling and in analysis runs performed on real data in the future.

\section*{Acknowledgments}
IH acknowledges and thanks the Max Planck Gesellschaft for support.
TH gratefully acknowledges support from the Radboud Excellence Initiative, from the International Centre for Theoretical
Sciences (ICTS) during a visit for participating in the program Summer School on Gravitational-Wave Astronomy
(Code: ICTS/Prog-GWS/2017/07), and from \emph{NewCompStar}, COST Action MP1304.
Simulations for this manuscript were conducted on the ``vulcan'' computing cluster at the Max Planck Institute
for Gravitational Physics in Potsdam-Golm, which is funded by the Max Planck Gesellschaft.
The authors would like to thank Alessandra Buonanno, Ben Lackey and Vivien Raymond for useful discussions about this manuscript.
The authors also thank Leslie Wade for reading through the manuscript and providing useful
feedback and comments.
\section*{References}

\bibliography{references}

\end{document}